\documentclass[%
 aip,
 amsmath,amssymb,
 reprint,%
]{revtex4-1}

\usepackage{graphicx}
\usepackage{dcolumn}
\usepackage{bm}

\usepackage[utf8]{inputenc}
\usepackage[T1]{fontenc}
\usepackage{mathptmx}
\usepackage{etoolbox}

\usepackage[english]{babel}

\usepackage{dcolumn}
\usepackage{amsmath}
\usepackage{amssymb}
\usepackage{bm}
\usepackage{graphicx}
\usepackage[export]{adjustbox}
\usepackage[normalem]{ulem}  
\usepackage[percent]{overpic}

\usepackage[usenames,dvipsnames]{color}
\definecolor{darkblue}{rgb}{0,0,0.6}
\definecolor{darkred}{rgb}{0.6,0,0}
\usepackage{hyperref}
\hypersetup{
colorlinks=true,			
urlcolor=darkblue,
citecolor=darkblue,		
linkcolor=darkblue,	
}
\usepackage{pict2e}

\definecolor{ao(english)}{rgb}{0.0, 0.5, 0.0}
\newcommand{\so}[1]{}      
\newcommand{\red}[1]{#1}      

\newcommand{\x}{_{\underline{x}}}

\newcommand{\gmax}{\Gamma}

\newcommand{\Einit}{E_{\text{init}}}
\newcommand{\Eliq}{E_{\text{liq}}}

\makeatletter
\def\@email#1#2{%
 \endgroup
 \patchcmd{\titleblock@produce}
  {\frontmatter@RRAPformat}
  {\frontmatter@RRAPformat{\produce@RRAP{*#1\href{mailto:#2}{#2}}}\frontmatter@RRAPformat}
  {}{}
}%
\makeatother
\begin{document}

\preprint{AIP/123-QED}

\title[]{The Fate of Shear-Oscillated Amorphous Solids}
\author{Chen Liu}
\affiliation{Laboratoire de Physique de l'Ecole Normale Sup\'erieure, Paris, France}

\author{Ezequiel E. Ferrero}
\affiliation{Instituto de Nanociencia y Nanotecnolog\'{\i}a, CNEA--CONICET, 
Centro At\'omico Bariloche, R8402AGP S. C. de Bariloche, R\'{\i}o Negro, Argentina.}

\author{Eduardo A. Jagla}
\affiliation{Centro At\'omico Bariloche, Instituto Balseiro, 
CNEA, CONICET, UNCUYO, R8402AGP S. C. de Bariloche, R\'io Negro, Argentina}

\author{Kirsten Martens}
\affiliation{Univ. Grenoble Alpes, CNRS, LIPhy, 38000 Grenoble, France
}

\author{Alberto Rosso}
\affiliation{Universit\'e Paris-Saclay, LPTMS, CNRS, 91405 Orsay, France}

\author{Laurent Talon}
\affiliation{Universit\'e Paris-Saclay, FAST, CNRS, 91405 Orsay, France}

\email{cl4175@columbia.edu, ferrero@cab.cnea.gov.ar}

\date{\today}

\begin{abstract}
The behavior of shear-oscillated amorphous materials is studied using a coarse-grained model.
Samples are prepared at different degrees of annealing and then subject to athermal and
quasistatic oscillatory deformations at various fixed amplitudes.
The steady-state reached after several oscillations is fully determined by the initial 
preparation and the oscillation amplitude, as seen from stroboscopic stress and energy 
measurements.
Under small oscillations, poorly annealed materials display {\em shear-annealing}, 
while ultra-stabilized materials are insensitive to them.
Yet, beyond a critical oscillation amplitude, both kind of materials display a {\it discontinuous} transition to the same {\it mixed} state  composed by a fluid shear-band  embedded in a marginal solid. 
Quantitative relations between  uniform shear and the steady-state reached with this protocol 
are established. 
The transient regime characterizing the growth and the motion of the shear band is also studied.
\end{abstract}

\maketitle

\section{Introduction}
Amorphous solids are a vast class of materials,
common in nature and ubiquitous for human applications.
Their mechanical behaviour is strongly affected by the preparation protocol:
poorly annealed materials, such as emulsions, foams or gels, are soft and ductile, 
while well annealed materials such as metallic glasses, ceramics and silica are hard 
and brittle\cite{BonnRMP2017,NicolasRMP2018}. 
Under a uniform deformation, the former tend to melt into a liquid, while 
the latter fail with a sharp stress-drop and the appearance of a thin liquid
shear-band\so{\cite{OzawaPNAS2018,PopovicPRE2018,BarlowPRL2020}}\red{~\cite{shi2006atomic,OzawaPNAS2018,PopovicPRE2018,BarlowPRL2020}}.
Although we tend to distinguish between these two kind of yielding,
a debate is open: ductile materials could also display stress 
overshoots\cite{BarlowPRL2020,ozawa2021rare, david2021finitesize,fielding2021yielding}.
Beyond finite-size issues\cite{ozawa2021rare,david2021finitesize, fielding2021yielding},
dealing only with transient states does not help to settle the discussion.
This is where an oscillatory deformation protocol\cite{Fiocco2014encoding,LeishangthemNC2017,das2018annealing,HexnerPNAS2020,sastry2020models,khirallah2021yielding,YehPRL2020,KawasakiPRE2016,RegevPRE2013,RegevNC2015} 
becomes handy, since it opens the possibility of characterizing a transition in terms
of stationary states.
In molecular dynamics (MD) simulations, two different ``phases'' have been identified: 
For moderate strain amplitudes $\Gamma$, the material appears solid and is progressively 
annealed\cite{LeishangthemNC2017,das2018annealing,YehPRL2020,RegevPRE2013,KawasakiPRE2016} 
as oscillation cycles accumulate.
Above a critical amplitude $\Gamma_c$,  further shear-annealing is prevented and a 
shear-band coexists with a solid\cite{fiocco2013oscillatory,RegevPRE2013,KawasakiPRE2016,LeishangthemNC2017,das2018annealing}.
Despite a rapidly growing literature\cite{fiocco2013oscillatory,RegevPRE2013,Fiocco2014encoding,RegevNC2015,KawasakiPRE2016,PriezjevPRE2016,priezjev2017collective,LeishangthemNC2017,regev2017irreversibility,regev2018critical,das2018annealing,MunganPRL2019,bhaumik2019role,ParmarPRX2019,das2020unified,YehPRL2020,HexnerPNAS2020,sastry2020models,khirallah2021yielding}, 
basic questions remain to be addressed: 
(i) the nature of the transition at $\Gamma_c$, 
(ii) the relation between steady-states in 
the two protocols (oscillatory and uniform shear), and
(iii) how those states are reached.

To answer these questions, in this manuscript we use a coarse-grained approach,
which goes beyond natural limitations of MD simulations.
We find that the transition from solid to flow at $\Gamma_c$ is discontinuous.
We quantitatively relate properties of the emerging phases in oscillatory shear with 
those of the uniform-deformation protocol.
Moreover, we describe transient stages previously unexplored.

To do that, we prepare samples at different degrees of annealing, characterized by the initial energy $\Einit$ per unit volume, and subject them to oscillatory deformations at various fixed amplitudes $\Gamma$ until steady-states are reached.
We find that once a {\it driving condition} ($\Einit$,$\Gamma$) is chosen, it univocally defines the material's fate in the steady-state; allowing us to fill a sort of {\it phase diagram}. 
Very well-annealed samples ($\Einit\!<\!E^*$, with $E^*$ a critical annealing level) 
are insensitive to small oscillations, while poorly-annealed samples ($\Einit\!>\!E^*$) 
exhibit {\it shear-annealing} for large enough sub-critical $\Gamma$.
We therefore distinguish between {\it stable solids}, that keep memory of
their initial condition, and {\it marginal solids} which result from poorly-annealed 
samples loosing memory of their initial conditions due to the shear-annealing by oscillations.
Yet, when the oscillation amplitude reaches the critical threshold 
$\Gamma_c(\Einit)$, every sample undergoes a \emph{discontinuous} transition, 
comprising a jump in stress and energy, towards {\it the same} mixed solid-fluid 
state.
Part of the system melt in a shear-band while the rest becomes  a critical solid 
of energy $E^*$, independently on the initial 
condition.
The fluid in the band corresponds to the one of the stationary state 
at large uniform quasi-static shear and holds at most the uniform yield 
stress $\Sigma_y$.
As $\Gamma$ is further increased, the width 
of the shear-band scales as a power of $(\Gamma-\Sigma_y/\mathcal{G})$,
with $\mathcal{G}$ the effective shear-modulus, further relating uniform 
and oscillatory protocols. 
Also, a rich transient dynamics towards the steady-state above $\Gamma_c$
is unveiled.
We show a band-width growth as a function of the number of oscillation 
cycles that is reminiscent of critical domain coarsening. 
Surprisingly, once the band reaches its stable width it ballistically sweeps out
the deeply-annealed regions, if present, turning them in the critical solid of 
energy $E^*$, and finally displays an (anomalous) diffusion.
Overall, our manuscript offers answers to fresh issues raised by MD simulations of glasses.

\section{method}
Our modelling is based on the evolution of a two-dimensional scalar strain field in 
a disordered potential, previously used to study yielding under uniform 
deformation\cite{JaglaPRE2007,FernandezAguirrePRE2018,FerreroPRL2019,JaglaPRE2020,FerreroJPCM2021}.
Initial sample annealing is achieved by tuning the configuration of the disordered landscape 
(lower initial energies $\Einit$ for better annealed systems).
Within this picture, the strain field behaves as a manifold propagating in 
a disorder medium\cite{LinPNAS2014,FernandezAguirrePRE2018,FerreroPRL2019}.
Two essential ingredients are incorporated:
the quadruple Eshelby's elastic interaction\cite{Eshelby1957,NicolasRMP2018} and
a disorder that rather than originating in quenched impurities 
mimics the random positions of particles.
The local disorder changes irreversibly, even if the 
strain manifold revisits the same location. 
Details of our method are presented in Appendix\ref{Apx-model}.

\section{Results}

\begin{figure}[t!]
    \centering
    \includegraphics[width=\columnwidth]{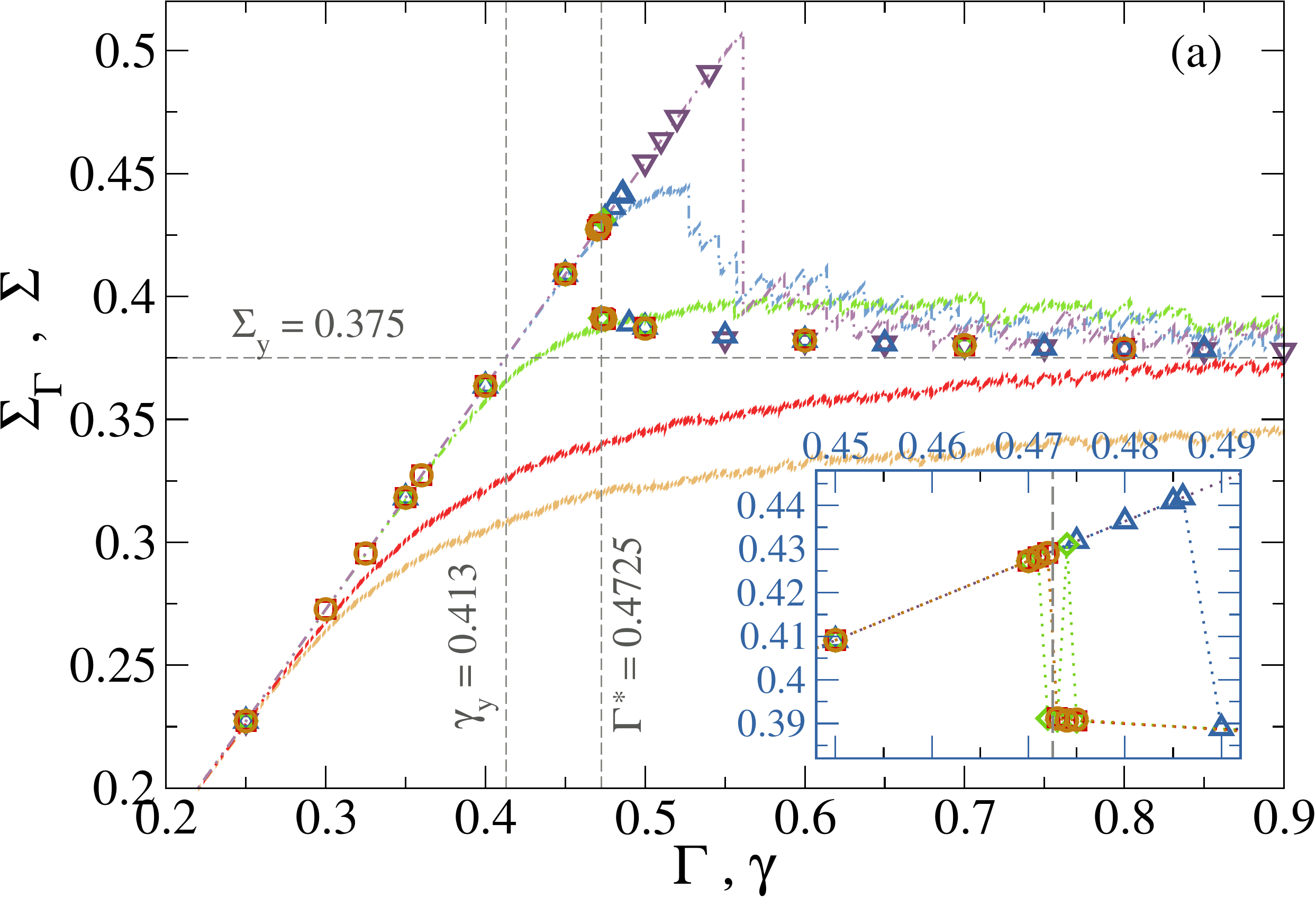}
    \mbox{ ~ }
    \includegraphics[width=\columnwidth]{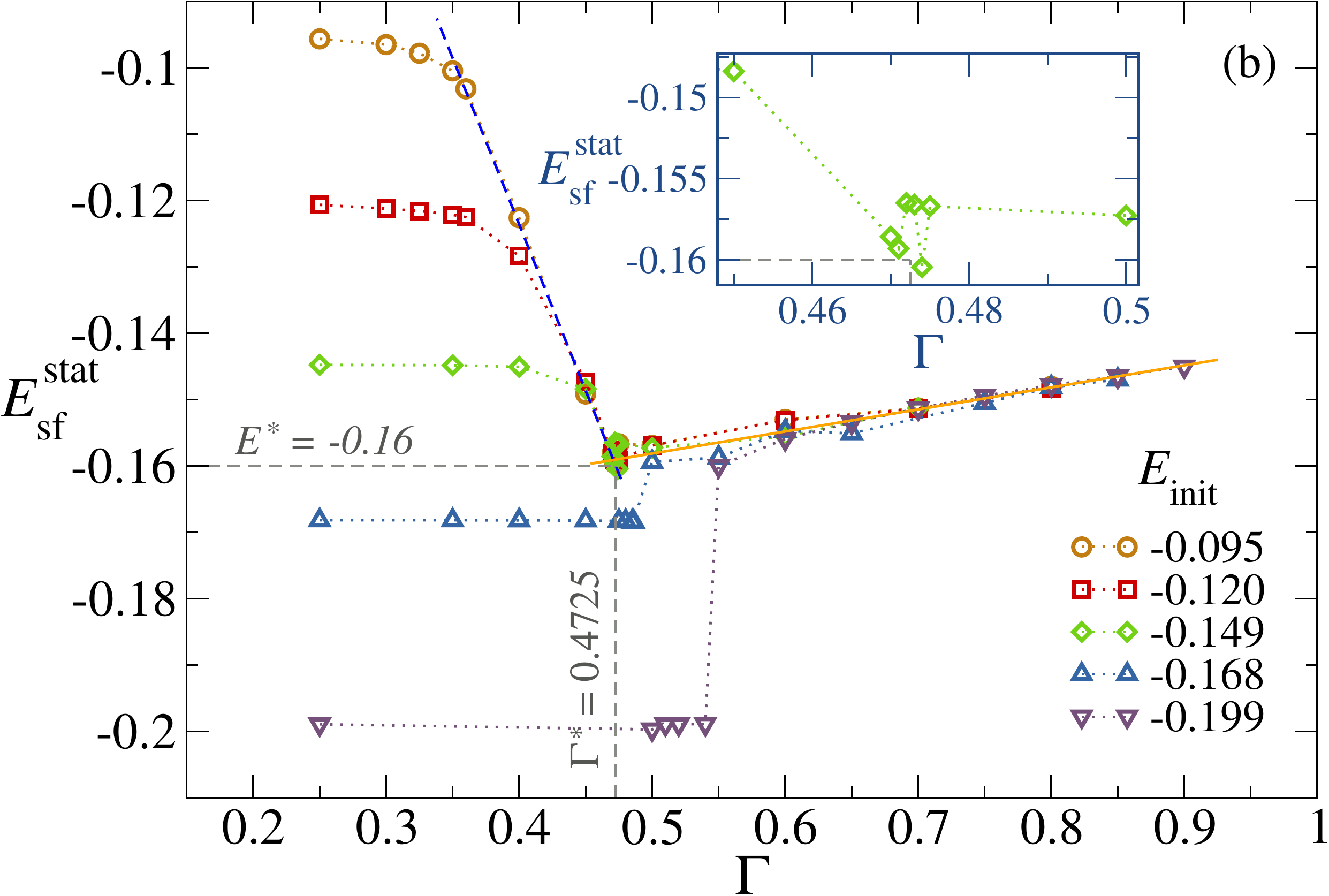}
    \caption{\emph{Stationary properties--} \textbf{(a)} 
    Symbols: Maximum stress \so{$\Sigma(\Gamma)$}\red{$\Sigma_\Gamma$} versus the oscillatory strain amplitude $\Gamma$. 
    Curves: Stress-strain \red{($\Sigma$ vs $\gamma$)} produced by quasi-static uniform shear. Different colors code for different degrees of initial annealing, signaled in the bottom panel. 
    \red{The inset shows a zoom-in on the transition region.}
    System size is $128^2$. 
    \textbf{(b)} Stress-free steady energy $E_{\tt sf}^{\tt stat}$ versus the $\Gamma$. The purple dashed-line corresponds to ``marginal solids''. The orange solid line displays a fit of $E_{\tt sf}^{\tt stat}(\Gamma)$ discussed in the text.
    \red{The inset shows a zoom-in on the transition region (only $\Einit=-0.149$ is shown).}
    }\label{fig:ssObserv_vs_gmax}
\end{figure}

In view of their relevance in interpreting the shear oscillation results,
we first reproduce some key features in uniform shear.
In Fig.\ref{fig:ssObserv_vs_gmax}(a) we show stress-strain curves for 
individual samples prepared at different degrees of annealing $\Einit$
(see also Appendix\ref{Apx-uni}).
After an initial linear elastic response of slope $\mathcal{G}\approx 0.91$, 
all samples reach at large strain $\gamma \gtrsim 3$ a common plateau 
corresponding to the  yield stress $\Sigma_y = 0.375\pm 0.003$.
The characteristic strain $\gamma_y$ where plasticity becomes important 
is estimated as $ \gamma_y = \Sigma_y/\mathcal{G} \simeq 0.413$. 
Poorly annealed materials display a monotonic crossover from an elastic 
solid to a liquid crossing $\gamma_y$, while well annealed systems remain 
elastic above $\gamma_y$ up to a failure strain $\gamma_{\tt f}$ where a sharp 
stress downfall occurs. 
The overshoot gets larger for better annealed samples\cite{OzawaPNAS2018,PopovicPRE2018,BarlowPRL2020,fielding2021yielding}.
In general, the total energy per volume of the material can be decomposed into two contributions: 
$E_{\rm tot} = E_{\tt sf} + \Sigma^2/(4\mu)$, 
where the second contribution is due to macroscopic elastic deformation 
(with $\mu=1/2$ the shear modulus) and $E_{\tt sf}$, called `stress-free energy' afterward, characterizes the state of the material with zero macroscopic stress.
In the large uniform shear limit $\gamma\rightarrow\infty$, where the memory of 
the initial state is completely erased, all samples converge to a homogeneous stationary 
liquid state with $E_{\tt sf}^{\tt stat} = -0.122\pm0.001\equiv\Eliq$ 
(see Fig.\ref{fig:uniformdeformation} in Appendix\ref{Apx-uni}). 

\subsection{Steady-state properties}

Now, in our oscillatory protocol, the material is deformed
by quasi-statically varying the strain from $0$ to $\Gamma$,
then to $-\Gamma$ and subsequently back to $0$ to complete 
one cycle of amplitude $\Gamma$.
This cycle is repeated until a steady-state is reached. 
By varying $\Gamma$ and $\Einit$, we obtain an assortment of stationary states.
Figure~\ref{fig:ssObserv_vs_gmax}(a) shows the steady-state 
maximum stress\cite{LeishangthemNC2017,fiocco2013oscillatory,YehPRL2020} 
\so{$\Sigma(\Gamma)$}\red{$\Sigma_\Gamma = \frac12 (\Sigma(-\Gamma)+\Sigma(+\Gamma))$} 
as a function of $\Gamma$ in comparison with uniform-shear load curves.
\red{The symmetric definition of $\Sigma_\Gamma$ is motivated in the need of 
suppressing the bias caused by the abundant plasticity in 
the initial direction of the shear.} 
Consistent with MD results, all samples exhibit a sharp jump of 
\so{$\Sigma(\Gamma)$}\red{$\Sigma_\Gamma$} at a critical strain amplitude $\Gamma_c(\Einit)$. 
The moderately annealed samples ($\Einit = -0.095$, $-0.12$ and $-0.149$) 
fail at the same $\Gamma_c \approx 0.4725 \equiv \Gamma^*$, 
independently on the initial degree of annealing; while the more 
annealed samples ($\Einit = -0.168$ and $-0.199$) experience a later
and stronger failure at $\Gamma_c(\Einit) \approx \gamma_{\tt f}(\Einit) >\Gamma^*$,
as better annealed is the sample. 
Once above $\Gamma_c$, the steady-state stress $\Sigma(\Gamma>\Gamma_c)$ for 
any initial state, is well identified with $\Sigma_y$ of a stationary flowing state in the quasi-static
uniform shear deformation.
Figure~\ref{fig:ssObserv_vs_gmax}(b) displays the stationary stress-free
energy $E_{\tt sf}^{\tt stat}$ as a function of $\Gamma$. 
\so{Here, u}\red{U}sing the stress-free energy $E_{\tt sf}$ has the advantage to
characterize the intrinsic state without external load, while the stroboscopic 
energy\cite{ParmarPRX2019,LeishangthemNC2017,das2020unified} incorporates an 
arbitrary elastic contribution when  $\Gamma>\Gamma_c$.
For very small $\Gamma$, the stationary state $E_{\tt sf}^{\tt stat}$ 
strongly depends on $\Einit$. 
Increasing $\Gamma$ within the solid phase ($\Gamma<\Gamma_c$), two scenarios 
separated by a critical annealing $E^*\approx -0.16$ are observed.
Samples with $\Einit < E^*$ keep a perfect memory of their initial states 
up to melting, evidenced by 
$E_{\tt sf}^{\tt stat}(\Gamma)=\Einit$ for all $\Gamma< \Gamma_c(\Einit)$ \cite{bhaumik2019role,YehPRL2020}.
Yet, for $\Einit > E^*$ the phenomenology is richer:
Increasing $\Gamma$, at some point, the stationary-state $E_{\tt sf}^{\tt stat}$ starts to decrease with the amplitude, loosing memory of the initial condition.
This is a manifestation of shear-annealing in the steady-state, as similarly observed in\cite{LeishangthemNC2017,ParmarPRX2019,das2020unified,bhaumik2019role,YehPRL2020}. 
Notice that, for initially weakly annealed systems, there exist also a {\it transient} 
process of shear-annealing (described later on), where samples reach a lower-energy
steady-state by oscillating at a fixed $\Gamma$.
$E_{\tt sf}^{\tt stat}$ for $\Einit>E^*$ ends up collapsing to a common curve 
(purple dashed line in Fig.~\ref{fig:ssObserv_vs_gmax}(b)),
where the initial condition becomes irrelevant and $E_{\tt sf}^{\tt stat}$ depends only 
on $\Gamma$.
We call ``marginal solids'' the stationary states lying on the purple dashed line,
which terminates at $\Gamma^*$ at the critical energy $E^*$.
Once $\Gamma>\Gamma_c(\Einit)$, both \so{$\Sigma(\Gamma)$}\red{$\Sigma_\Gamma$} and 
$E_{\tt sf}^{\tt stat}(\Gamma)$ fall onto a unique curve, independent on $\Einit$. 
The system develops a permanent liquid shear-band embedded in a solid phase, 
and the steady-state energy $E_{\tt sf}^{\tt stat}$ increases with $\Gamma$
(Fig.~\ref{fig:ssObserv_vs_gmax}(b)), see also\cite{ParmarPRX2019,fiocco2013oscillatory,YehPRL2020}.
From the previous discussion, we can associate to each $\Einit$ an amplitude 
$\Gamma_a(\Einit)$ above which the steady-state $E_{\tt sf}^{\tt stat}(\Gamma,\Einit)$ is independent on the initial annealing $\Einit$. 

\begin{figure}[t!]
    \centering
    \includegraphics[width=0.9\columnwidth]{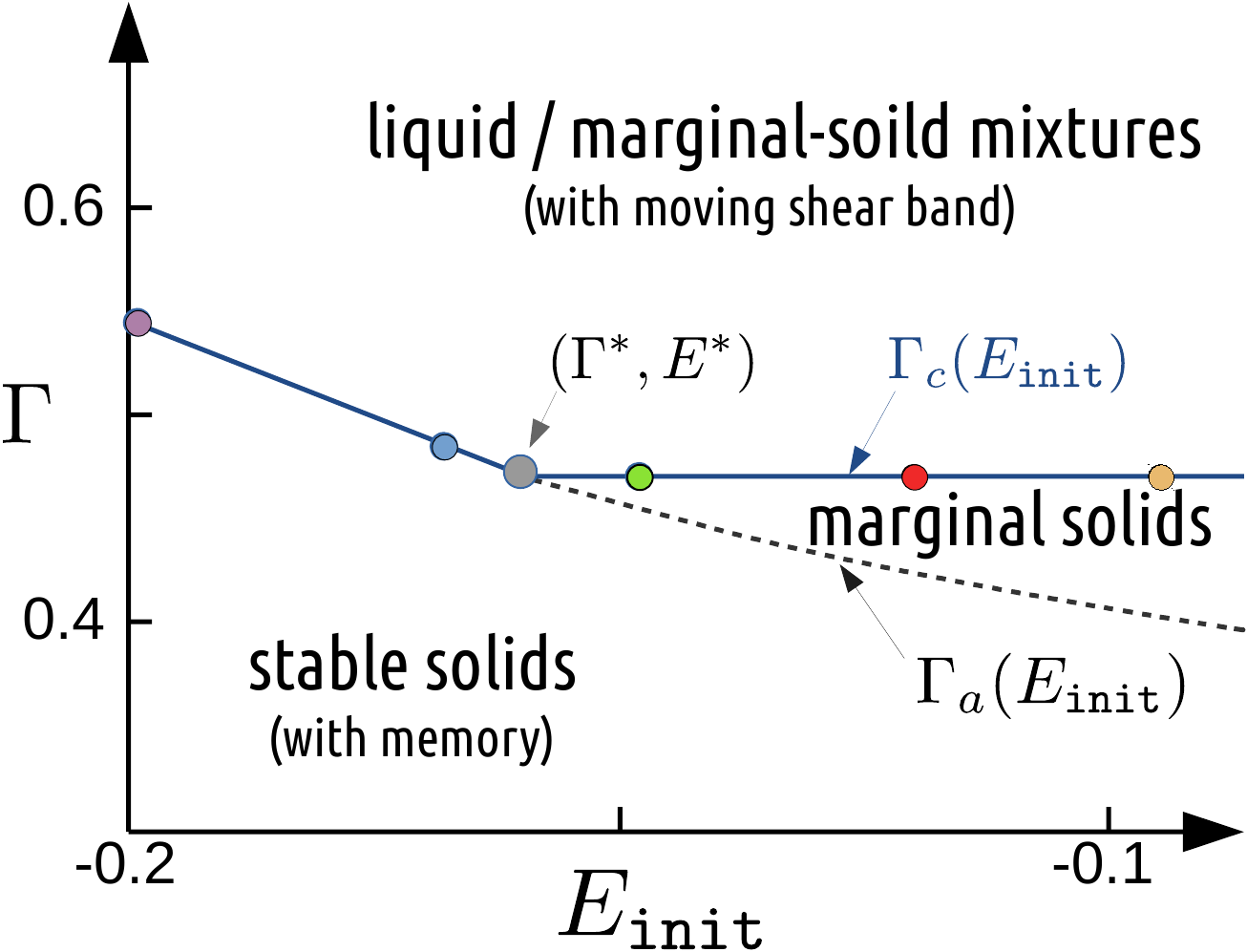}
    \caption{``Phase diagram'' classifying driving conditions ($\Einit$,$\Gamma$) according to the steady-state they will reach: \emph{stable solids} (strongly reminiscent of the initial state), \emph{marginal solids}  (shear-annealed solids independent of the initial condition), and \emph{solid-liquid mixtures} characterized by an erratic shear-band. Color circles indicate $\Gamma_c$ for different initial degrees of annealing $\Einit$.}
    \label{fig:phase}
\end{figure}

As summarized in a diagram Fig.\ref{fig:phase}, 
$\Gamma_a(\Einit)$ (dashed line) identifies with $\Gamma_c(\Einit)$ 
(solid line) below $E^*$, and bifurcates above it. 
The solid steady-states resulting from driving conditions below $\Gamma_a(\Einit)$
are strongly reminiscent of the initial state and called ``stable solids''.
In contrast, those driving conditions in between $\Gamma_c$ and $\Gamma_a$ 
correspond to solids that forget their initial condition and end up being
``marginal solids'' in the steady-state.
Driving conditions in the region above $\Gamma_c(\Einit)$ lead to a steady-state 
mixed phase composed by a fluid shear band surrounded by a critical marginal solid, 
which we detail in the following.

\begin{figure}[t!]
    \centering
    \includegraphics[width=1\columnwidth]{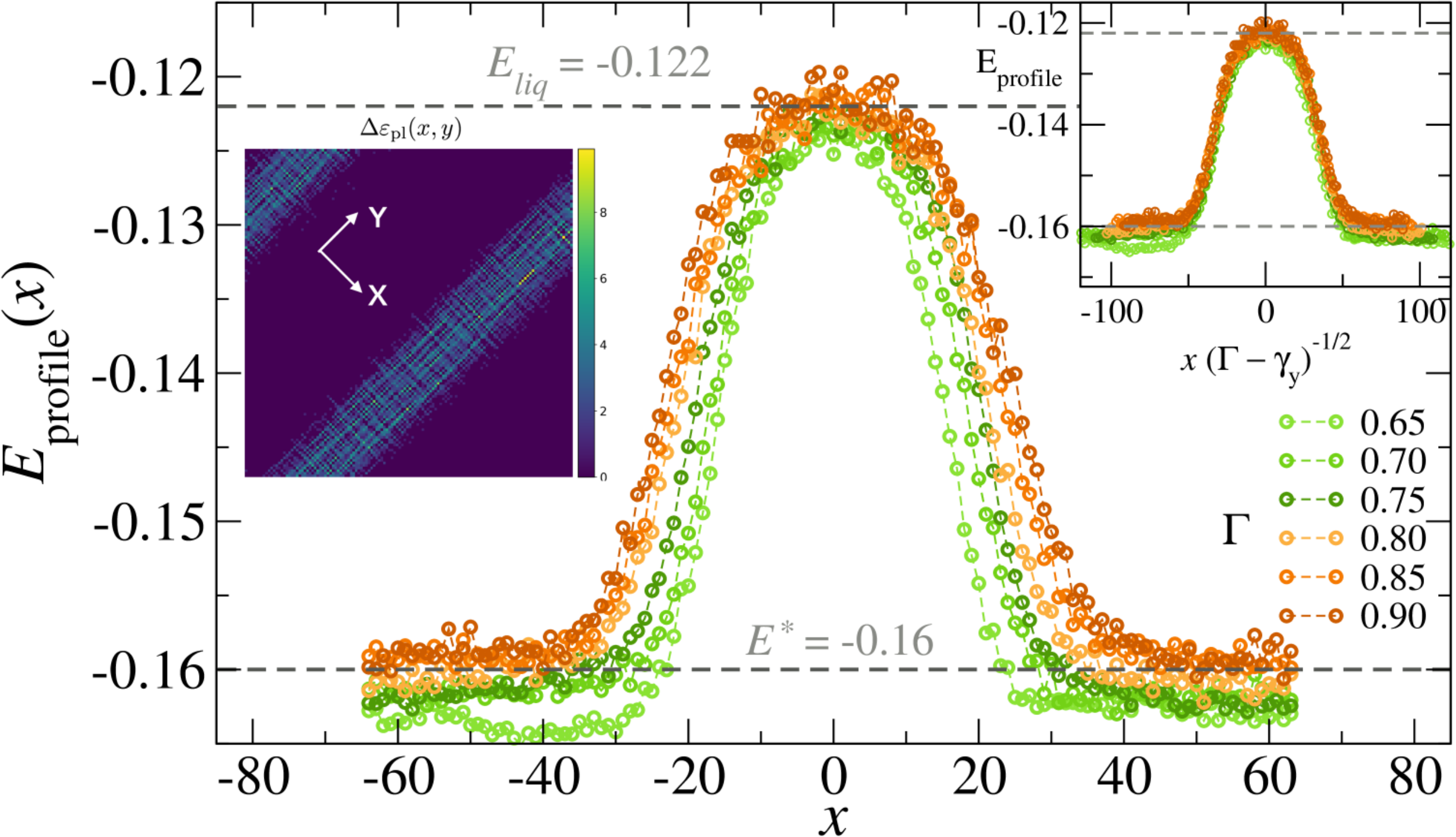}
    \caption{Energy profiles averaged along the y-axis for different $\Gamma$. $x=0$ is set to the maximum location. {\it Left inset:} map of strain field during a half-cycle in the steady-state. {\it Right inset:} profiles collapse using $w_s(\Gamma) = a ( \Gamma - \Gamma_0)^\alpha$ with $\Gamma_0=\gamma_y=0.413$ and $\alpha=1/2$.}
    \label{fig:mixture}
\end{figure}

Fig.~\ref{fig:mixture} shows the averaged stress-free energy profile across a section 
orthogonal to the shear-band\red{~\cite{manning2007strain}} 
(evidenced in the strain field in the left-inset).
The energy profile characterizes the band by a bell-shape. 
The interior of the band (top of the bell) has the same energy as the
stationary liquid in uniform shear deformation $\Eliq\approx-0.122$.
The width $w_s$ of the band increases with $\Gamma$ and is well 
fitted by $w_s(\Gamma) = a ( \Gamma - \Gamma_0)^\alpha$.
A good collapse is found when rescaling the profiles with $\alpha=1/2$
(right-inset of Fig.~\ref{fig:mixture}), reminiscent of the transient 
band dynamics in uniform shear\cite{JaglaJSTAT2010}. %
We further notice two peculiarities: 
First, from the fit, $\Gamma_0$ identifies with $\gamma_y$ instead of $\Gamma^*$, 
implying that at the transition, the shear-band 
has a finite width $w_s(\Gamma^*)>0$ (because $\gamma_y<\Gamma^*$).
Second, the energy profile is independent on
$\Einit$ and the energy at the bottom of the bell-shape coincides with the critical 
solid energy $E^*$. 
This allows to conclude that the mixture phase is completely independent on the 
initial degree of annealing and uniquely characterized by the width $w_s$: 
the stationary regime is composed of a fraction of critical solid at energy $E^*$ 
and a shear-band made of the liquid of energy $E_{\rm liq}$.  
Since $\Eliq>E^*$, the monotonic growth of $E_{\tt sf}^{\tt stat}$ (Fig.~\ref{fig:ssObserv_vs_gmax}(b)) 
can be rationalized in terms of the shear-band widening with $\Gamma$:
\so{$E_{\tt sf}^{\tt stat} \simeq E^* + \left(E_{\rm{liq}} - E^*\right)\frac{a(\Gamma - \gamma_y)^{\alpha}}{L}. $}
\red{
\begin{equation}\label{eq:esf}
    E_{\tt sf}^{\tt stat} \simeq E^* + \left(E_{\rm{liq}} - E^*\right)\frac{a(\Gamma - \gamma_y)^{\alpha}}{L}. 
\end{equation}
}

\so{This} \red{The validity of this behavior} is tested in
Fig.~\ref{fig:ssObserv_vs_gmax}(b) \red{(orange line)}.
\red{
From our analysis, the transition seems to be of first order in nature 
for two reasons:
First, the jump in $E_{\tt sf}^{\tt stat}$ is clearly discrete
for well annealed samples, but it appears discontinuous also for the poorly 
annealed systems, where around $\Gamma^*$ we observe energy fluctuations 
between two close but distinct levels (see the inset of Fig.~\ref{fig:ssObserv_vs_gmax}(b)):
$E_{\tt sf}^{\tt stat} \approx -0.16$ for the realizations 
that remain in the solid phase and $E_{\tt sf}^{\tt stat} \approx -0.157$
for those that transition to the mixed phase.
Second, from the finite-size analysis of $E_{\tt sf}^{\tt stat}$ 
showing a nice collapse of $E_{\tt sf}^{\tt stat}$ for different system
sizes $L$ in Fig.~\ref{fig:Esf_stat} (Appendix.\ref{Apx-dis}), we can conclude that the shear band 
when appearing invades a finite portion of the 
system~\cite{ParmarPRX2019,YehPRL2020,das2020unified,bhaumik2019role,LeishangthemNC2017} $w_s(\Gamma^*)/L>0$. The first-order nature of the transition in energy $E_{\tt sf}^{\tt stat}$ at $\Gamma^*$ is further elaborated in Appendix.\ref{Apx-dis}.
}

\so{
It also implies that a discontinuity of $E_{\tt sf}^{\tt stat}$ at the transition 
$\Gamma_c=\Gamma^*$, even though small, is expected\cite{ParmarPRX2019,YehPRL2020,das2020unified,bhaumik2019role,LeishangthemNC2017}. 
In addition, we observed that $w_s$ is proportional to the linear system size, as 
also seen in the fact that $E_{\tt sf}^{\tt stat}$ shows no size dependence~\cite{SM}
}

\subsection{Transient properties}

\begin{figure}[t!]
    \centering
    \includegraphics[width=0.96\columnwidth]{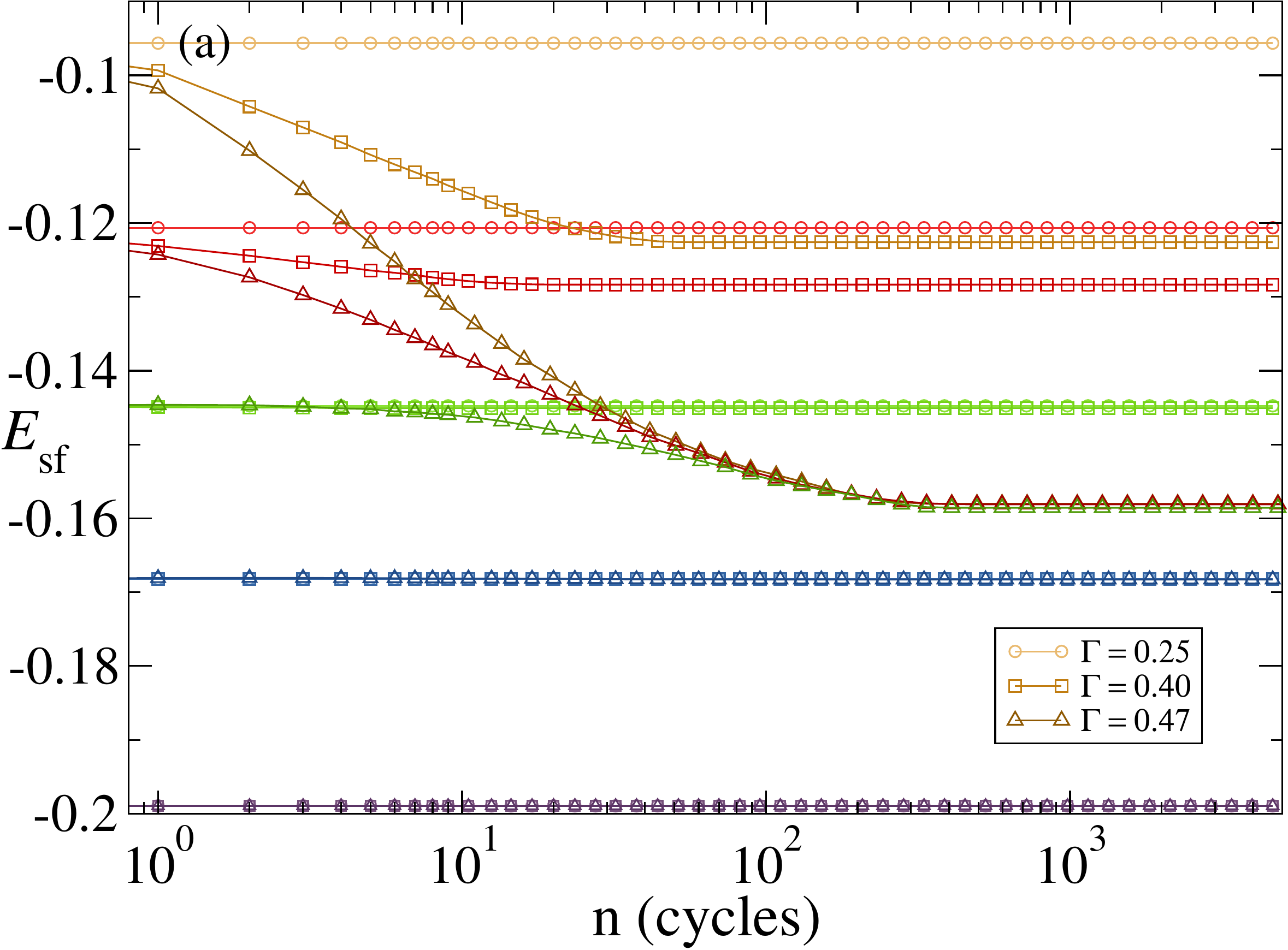}
    \mbox{ ~ }    
    \includegraphics[width=0.96\columnwidth]{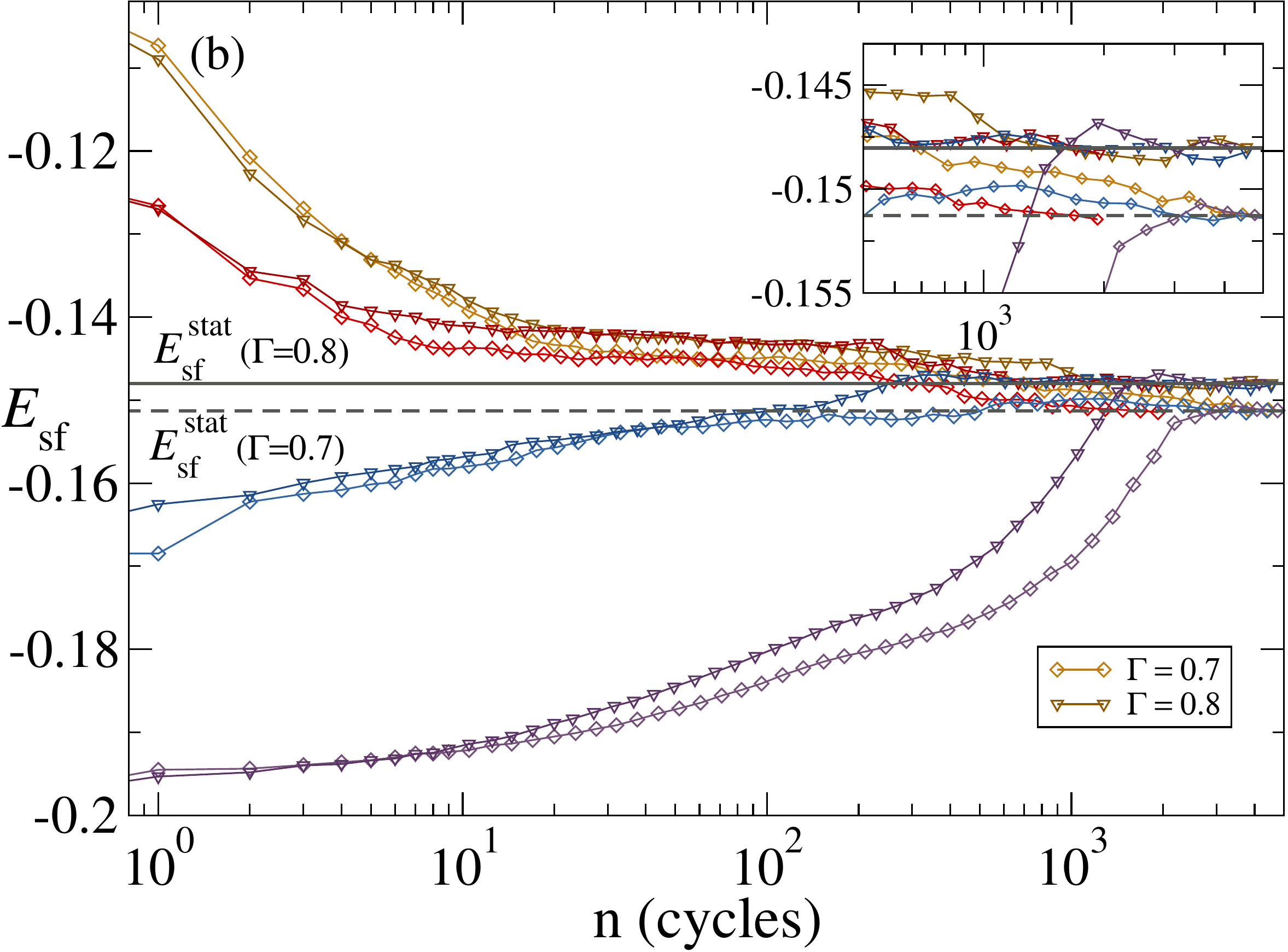}
    \caption{\emph{Transient properties--} Evolution of stress-free energy $E_{\tt sf}$ as the number of cycles $n$ increases, for different amplitudes. Colors coding for different $\Einit$ are the same as in Fig.\ref{fig:ssObserv_vs_gmax}. System size is $128^2$. \textbf{(a)} Solid phase ($\Gamma\!<\!\Gamma_c$). \textbf{(b)} Mixed phase ($\Gamma\!>\!\Gamma_c$). The inset shows a zoom-in to appreciate the difference in $E_{\tt sf}^{\tt stat}$ at different $\Gamma$.}
    \label{fig:transientEnergy}
\end{figure}

We now discuss the transient oscillatory dynamics. 
Fig.\ref{fig:transientEnergy} shows the evolution of $E_{\tt sf}(n)$, as
the oscillation cycles $n$ accumulate, for different driving conditions 
$(\Einit, \Gamma)$.
Below $\Gamma_c$ (Fig.\ref{fig:transientEnergy}(a)), ultra-stable systems 
($\Einit<E^*$) are unperturbed by the oscillations, while systems with
$\Einit>E^*$ shear-anneals if the amplitude is large enough. 
At amplitudes above $\Gamma_a(\Einit)$ all initial conditions
are shear-annealed to the same stationary state, as observed 
for our three softest samples in Fig.~\ref{fig:transientEnergy}(a).
Above $\Gamma_c$ (Fig.\ref{fig:transientEnergy}(b)), all samples
go to the same stationary state at large $n$.
Our observations for the number $n_T$ of cycles to reach steady-states are compatible with 
a divergence when approaching $\Gamma_c$ reported in previous works\cite{LeishangthemNC2017,khirallah2021yielding,sastry2020models,fiocco2013oscillatory,KawasakiPRE2016}. 
Data for ($\Gamma<\Gamma_c$) is shown in Appendix\ref{Apx-nt}.

\begin{figure}[t!]
\centering
\includegraphics[width=\columnwidth]{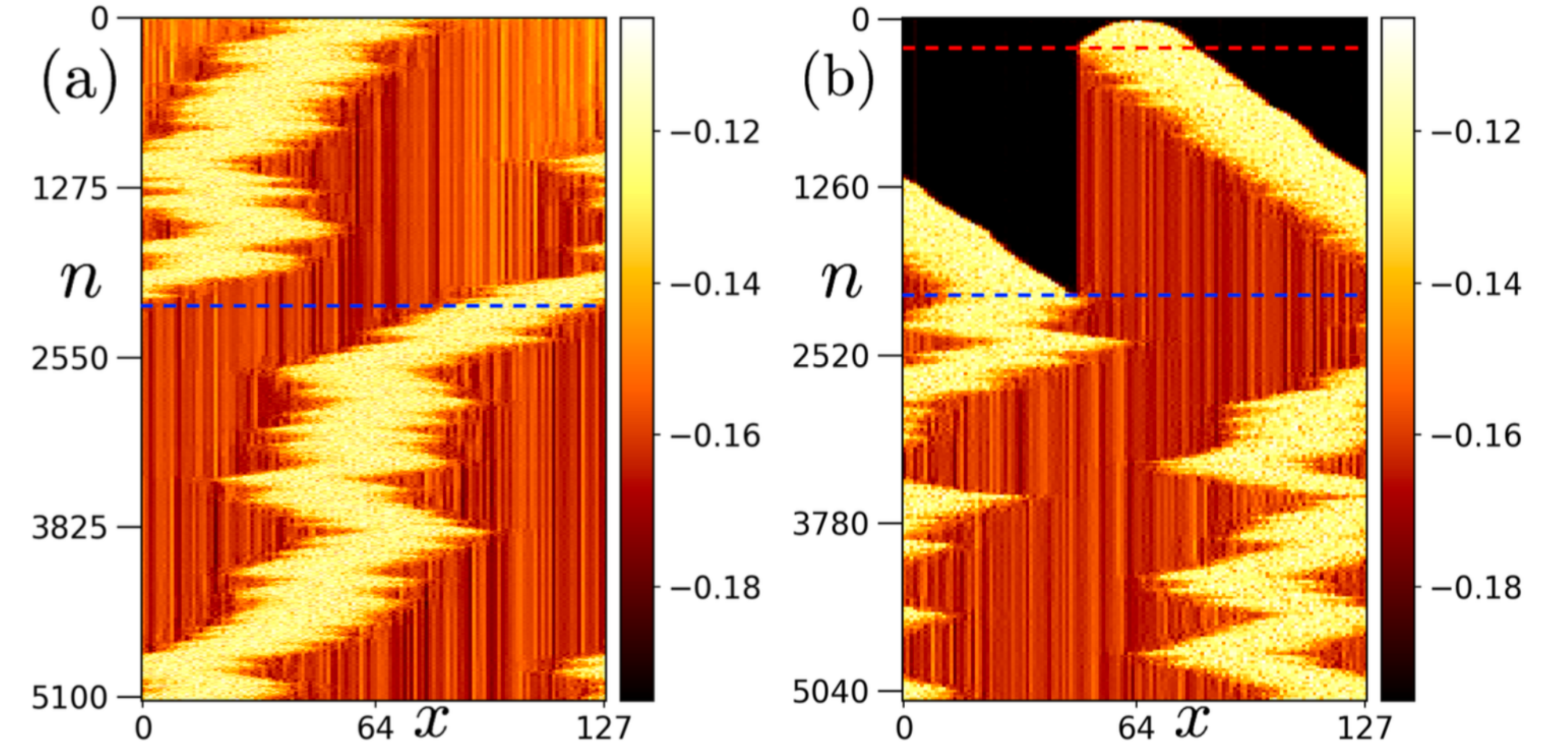}
\caption{Energy profiles $E_{\tt profile}(x)$ evolving as function of the number $n$ of oscillation cycles for $\Gamma=0.7$. Light colors represent higher energies and evidence the position of the shear-band. {\bf (a)} A poorly annealed sample with $\Einit\approx -0.12$ (see also Video 1 (Multimedia view)). {\bf (b)} Well annealed sample with  $\Einit\approx -0.199$ (see also Video 2 (Multimedia view)).  The red dashed-line in (b) indicates the stabilization of the band width at $n\approx 220$. When the blue dashed-line is reached, the band has visited the entire system.}
\label{fig:Prfl_evo}
\end{figure}

\begin{figure}[b!]
\centering
\includegraphics[width=0.95\columnwidth]{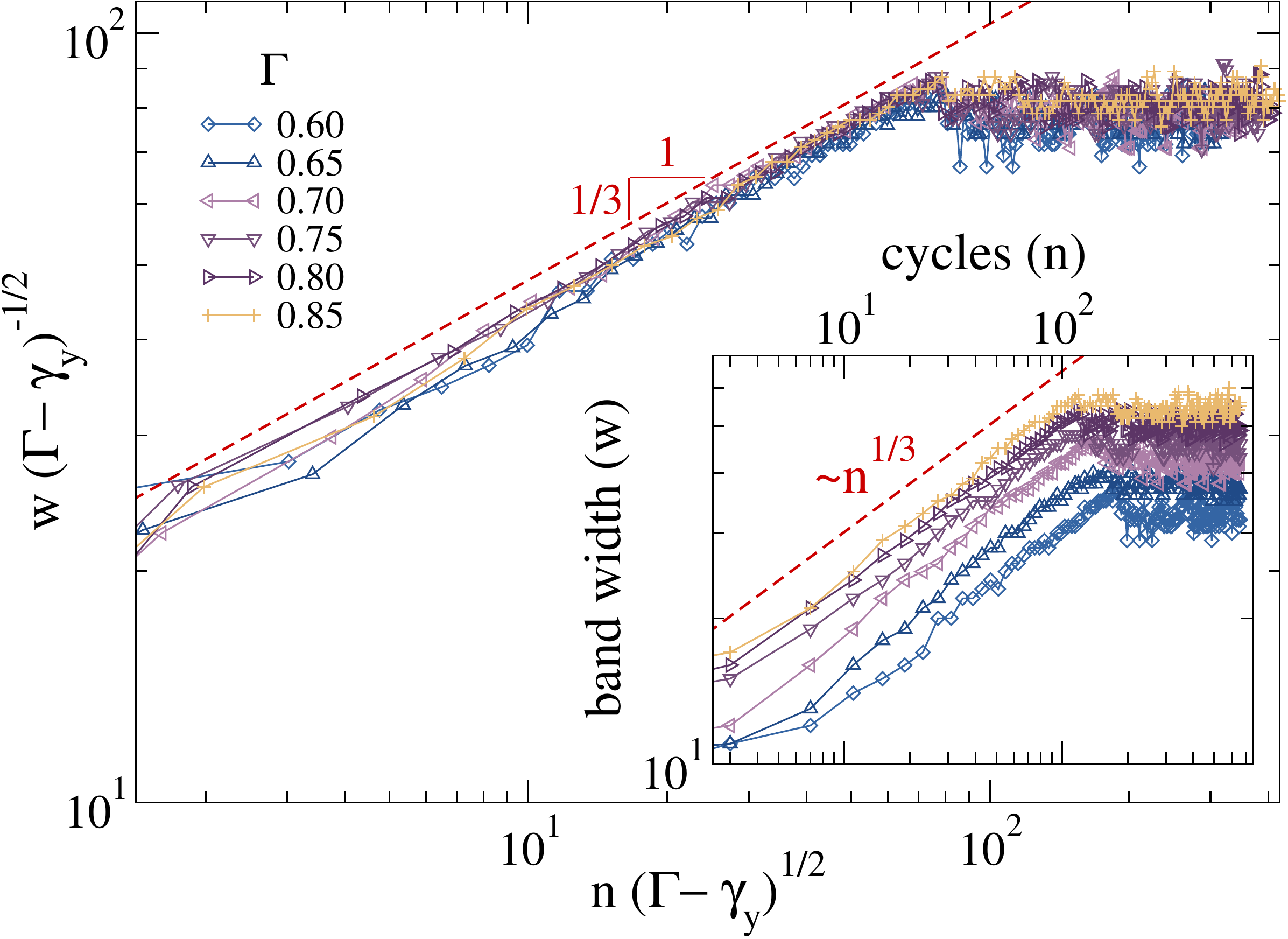}
\caption{Shear-band width $w$ as a function of $n$ for different $\Gamma$. The initial condition is a well annealed solid ($\Einit=-0.168$). The inset shows raw-data and the main panel the scaling $w(\Gamma-\gamma_y)^{-1/2}$ vs. $n(\Gamma-\gamma_y)^{1/2}$. The red dashed line is a power-law $\sim n^{1/3}$. System size is $256^2$.}
\label{fig:band_growth}
\end{figure}

We switch now to the observation of different dynamical regimes for the mixed phase.
For poorly annealed samples, the onset of the shear-band and its stabilization in 
a stationary width $w_s(\Gamma)$ occurs gradually and rapidly as the solid region
shear-anneals to the critical state in a small number of cycles. 
Then the shear band diffuses in the material (see Fig.~\ref{fig:Prfl_evo}(a) 
and Video 1 (Multimedia view)) with a mobility that increases with $\Gamma$.
In well annealed systems instead, the transient is richer and we identify 
three dynamical regimes (Fig.~\ref{fig:Prfl_evo}(b), see also Video 2 (Multimedia view)):
{\it (i) Shear band formation and growth}. 
The initial band growth is observed at the top of Fig.~\ref{fig:Prfl_evo}(b)
(see Fig.\ref{fig:Prfl_evo_Si} in Appendix\ref{Apx-growth} for a more detailed 
illustration for the initial band growth). 
This initial coarsening of the shear-band is studied in 
Fig.~\ref{fig:band_growth} at different $\Gamma$.
In all cases, the band width grows as $\sim n^{1/3}$ until reaching the stationary width
$w_s\sim(\Gamma-\gamma_y)^{1/2}$.
{\it (ii) Melting of the solid}. 
The now fully-formed shear band displays a ballistic motion preferentially invading 
the deeply annealed solid 
and leaving behind the critical solid of energy  $E^*$.
{\it (iii) Shear band diffusion}. 
When the shear-band has visited the entire system, all the regions outside 
the band have been modified into the critical marginal solid.
The stationary state is reached and the band diffuses
forever, maintaining its characteristic $\Gamma$-dependent width.

\section{Discussion} 
Our mesoscopic model allows to recover the phenomenology of the oscillatory shear of amorphous materials reported by MD simulations\cite{LeishangthemNC2017,bhaumik2019role,das2020unified,OzawaPNAS2018,YehPRL2020,das2020unified} and go beyond. 
While some emerging features can be already captured by even further simplified
toy models as seen recently\cite{sastry2020models}, keeping an spatial extent for 
the system and its geometry gives access to the full picture.
We are able to univocally classify the oscillatory stationary states according to 
a ``phase diagram'' in the space of oscillation amplitude $\Gamma$ and initial 
annealing level $\Einit$, and relate them with the steady-states of the 
uniform-deformation protocol.
Interestingly, the oscillatory protocol always shows a discontinuous transition 
in both stress and energy at a critical strain amplitude $\Gamma_c$,
even in the case of ductile materials that melt homogeneously under uniform 
deformation.
Notably, ductile materials mechanically anneal and harden at sub-critical 
amplitudes under oscillatory shear.
Independently on the initial state, every system sheared at large amplitudes 
ends up in a steady-state displaying a permanent shear-band.
This band contains a stationary fluid identical to the one obtained at large
uniform deformations and is embedded in a solid matrix which is not at all arbitrary.
The solid surrounding the shear-band in this mixed phase is the critical marginal solid
that has the energy of the critical annealing level $E^*$.
A line of {\it amnesia} $\Gamma_a(\Einit)$ discriminate driving conditions 
($\Einit$, $\Gamma$) between {\it stable solids} and
solids that shear-anneal and evolve towards a marginal line in the steady-state.
Our approach gives also transparent access to the transient regimes of shear-band
formation, growth and motion.
For deeply-annealed samples, a small but finite width band first spreads the system, 
then grows with a power-law $\propto n^{1/3}$ of the number of oscillation cycles.
Ulteriorly, the band invades ballistically the deeply-annealed solid, leaving behind 
the critical solid of energy $E^*$.
Finally, is diffuses anomalously over all the system.

These results shed light to clarify some previous observations in the literature
and open interesting directions of new research. 
One could ask, for example, whether the emergence of a transient state of multiple 
shear-bands is possible in the large system limit, and how do they merge in a 
single band in the steady-state (or not).
Also, to which extent the incorporation of ``reversible'' plastic events 
observed in atomistic simulations\cite{MunganPRL2019,KawasakiPRE2016,RegevPRE2013,RegevNC2015,PriezjevPRE2016,regev2018critical, fiocco2013oscillatory} enriches the overall phenomenology. 
Those are absent in our present model by choice, but easy to include by locally quenching the disordered potential~\cite{khirallah2021yielding}.
Further, the critical annealing $E^*$ detected in the oscillatory protocol 
could play also a relevant role in uniform shear, where the ductile-brittle 
yielding transition is a matter of vivid discussion\cite{OzawaPNAS2018,OzawaPNAS2018,fielding2021yielding,BarlowPRL2020}.
It is indeed suggestive that the load curve of our closest to $E^*$ sample sits near to the 
apparent transition between ductile/brittle yielding probed by the uniform shear curves. 
More importantly, it will be interesting to test experimentally these predictions, 
with new measurements closer to the quasi-static limit, since in the standard 
setups the oscillations are relatively fast and the material softens instead of 
annealing and hardening. 

\red{Moreover, general features of our phase diagram could find analogies in
different systems with elastic interactions, irreversible events and quenched disorder,
such as vortex or skyrmion systems where the initial state can be also prepared in 
different ways~\cite{mangan2008reversible,okuma2011transition,brown2019reversible,maegochi2021critical}.
}
In general, we hope that our results will motivate new research directions in the study 
of low temperature amorphous materials, from realistic large-scale atomistic simulations 
to experiments.
\vspace{-0.5cm}


\begin{acknowledgments}
We are indebted to M. Ozawa for illuminating discussions on the subject.
We thank S. Sastry for kind feedback on an earlier version of this manuscript. 
We acknowledge support from the collaboration project ECOS Sud-MINCyT A16E01
and IRP project `Statistical Physics of Materials' funded by CNRS.
CL, AR and LT acknowledges support by “Investissement d’Avenir”
LabEx PALM (Grant No. ANR-10-LABX-0039-PALM)
\end{acknowledgments}

\appendix

\section{Methods: Our Model\label{Apx-model}}
We focus on two dimensional systems. 
Our goal is to provide a description of the physics of
amorphous materials under deformation and we use a model  based on the evolution of the three  strain field components~\cite{JaglaPRE2007,cao2018soft}: 
the isotropic compression $\varepsilon_1(\mathbf{r})$ and the shear components  $\varepsilon_2(\mathbf{r})$ and 
 $\varepsilon_3(\mathbf{r})$.
 The latter are bi-axial deformations (compression along some direction, expansion perpendicularly), the difference between $\varepsilon_2(\mathbf{r})$ and $\varepsilon_3(\mathbf{r})$ is that the compression/expansion axis are the $x-y$ axis for the first, and the lines at $45$ degrees for the second
(Recall that given the continuum displacement field,
$\mathbf{u(r)}=\left(u_1(\mathbf{r}),u_2(\mathbf{r})\right)$, the elements 
of strain tensor are given by $\epsilon_{ij}(\mathbf{r})=\frac{1}{2}(\partial_{i}u_{j}(\mathbf{r})+\partial_{j}u_{i}(\mathbf{r}))$. Then $\varepsilon_{1}(\mathbf{r})=\left(\epsilon_{11}(\mathbf{r})+\epsilon_{22}(\mathbf{r})\right)/2$, $\varepsilon_{2}(\mathbf{r})=\left(\epsilon_{11}(\mathbf{r})-\epsilon_{22}(\mathbf{r})\right)/2$ and $\varepsilon_{3}(\mathbf{r})=\epsilon_{12}(\mathbf{r})$). 
For a perfectly elastic solid, one can write the energy of any deformation in 
terms of a quadratic elastic functional 
\begin{equation}\label{eq:elasticfunctional}
    E_{\text{elast.}}=\int d^2 \mathbf{r}\, \left(B\varepsilon_1^2+\mu \varepsilon_2^2 +\mu \varepsilon_3^2 \right) \, ,
\end{equation}
where $B$ and $\mu$ are respectively the bulk and shear modulus. Here we use $B=1$ and $\mu=1/2$. 
This purely elastic functional has a single minimum, which corresponds to the undeformed state. 
When a deformation is applied, it brings the system out of such a minimum; 
when released, a restoring stress brings the system back to its undeformed state.
If we want to take plasticity into account, irreversible deformations should be allowed;
for example, the energy landscape may be characterized by multiple local minima separated by finite barriers \red{(similar to the case of sheared vortices in random quenched potentials~\cite{maegochi2021critical, okuma2011transition})}.
Here, we study the response of the material under simple shear. 
We  account for the plastic events by replacing the harmonic term $\mu \varepsilon_3^2$ 
in Eq.~\ref{eq:elasticfunctional} with a disordered potential $V\x[\mathbf{r}, \varepsilon_3]$
that displays many minima as $\varepsilon_3$ varies. 
The other two strain components, $\varepsilon_1$ and $\varepsilon_2$, 
are expected to remain small, and the quadratic approximation, to hold.
The energy functional then becomes 
\begin{equation}\label{eq:e_plt}
    E_{plast.}=\int d^2 \mathbf{r} \, \left(B\varepsilon_1^2+\mu \varepsilon_2^2 +V\x[\mathbf{r}, \varepsilon_3] \right) \; 
\end{equation}
The subscript of $V\x$ represents an internal degree of freedom. 
Thus the disorder potential is not a function of the strain field $\varepsilon_3(\mathbf{r})$ only. 
The origin of that internal degree of freedom is justified as follows:
In amorphous materials, the disorder is not generated by immobile impurities 
but rather by the random-like configurations of the particles position $\underline{x}$.
Plastic events, called {\em shear transformations}, are  localized in space.
When an event occurs in the region $\mathbf{r}$, the particles inside the region  
rearrange irreversibly from $\underline{x}$ to $\underline{x}'$. 
This rearrangement is mimicked in the model by two features. 
\begin{itemize}
\item First, $\varepsilon_3(\mathbf{r})$ 
overcomes a local energetic barrier
and reach the basin of a new  minimum of the disordered potential.
\item Second, the basin of the old minimum also modifies irreversibly as new configuration of particles positions imposes a new  disorder potential.
\end{itemize}
In the spirit of keeping only the essence of the physical process 
involving local plastic events, we  model the disorder of each region $\bm{r}$
as a collection 
 of pieces of harmonic potential: 
 \begin{equation}\label{eq:local_potential}
 V_{\text{parab}} (\varepsilon_3) = 
 \mu[(\varepsilon_3-\epsilon_{pl})^2-\epsilon_{w}^2], \quad \text{for} \, |\varepsilon_3- \epsilon_{pl}|<\epsilon_w.
\end{equation}
The random parameters $\epsilon_{pl}$ and $\epsilon_w$ vary from region 
to region: 
$\epsilon_{pl}$ defines the location of the minimum and $\epsilon_w$, 
the semi-width of the potential. 
Once the strain $\varepsilon_3$ overcomes the barriers of the parabolic 
basin a new semi-width $\epsilon_w'$ is drawn from a triangular distribution 
of mean $1/2$ and standard-deviation $1/20$. 
The new  minimum of the potential is updated as follows: 
\begin{itemize}
    \item When  $\varepsilon_3$ overcomes the barrier at $\epsilon_{pl} -\epsilon_w$ then the new minimum is located at $ \epsilon_{pl} -(\epsilon_w+ \epsilon_w')$ 
    \item When  $\varepsilon_3$ overcomes the barrier at $\epsilon_{pl} +\epsilon_w$ then the new minimum is located at $ \epsilon_{pl} +(\epsilon_w+ \epsilon_w')$. 
\end{itemize}
It is important to stress that once strain $\varepsilon_3$ overcomes the barriers of the parabolic 
basin  the local basin changes irreversibly, even is the $\varepsilon_3$ moves backward.

The three components of the strain field are not independent but obey 
the St. Venant constraint~\cite{JaglaPRE2007}.
As a consequence, the deformation of the material can be described by the 
evolution of a single component. 
In particular assuming, in the framework of an overdamped dynamics, that $\varepsilon_1$ and $\varepsilon_2$ relax much faster than $\varepsilon_3$, 
we arrive to
\begin{equation}\label{eq:dym1}
    \partial_{t}\varepsilon(\mathbf{r},t)=\int d^2\mathbf{r}'G(\mathbf{r-r'})\varepsilon(\mathbf{r}')+\eta_{\x}(\mathbf{r},\varepsilon)+\Sigma_{\text{ext}}.
\end{equation} 
Here, $\varepsilon(\mathbf{r},t)$ has replaced $\varepsilon_3(\mathbf{r},t)$ 
for a lighter notation and the kernel $G(\mathbf{r})$ is the long-range Eshelby's 
propagator.
For simple shear, using the method in \cite{cao2018soft}, the propagator can be written in terms of the polar 
coordinates $\mathbf{r}=(r,\theta_{\mathbf{r}})$ as 
\begin{equation}
	G(\mathbf{r}) = \frac{2\sqrt{2}B}{\pi r^2}\frac{3\cos(4\theta_{\mathbf{r}})-1}{\left(3-\cos(4\theta_{\mathbf{r}}) \right)^2}\, .
\end{equation}
Finally, $\eta_{\x}$ is originated from the disorder potential as 
$-\partial_\varepsilon V_{\x}$ and $\Sigma_{\text{ext}}$ is the external applied stress. 

For our purposes, it is important to control the macroscopic strain instead of the 
macroscopic stress. 
In order to achieve this, one can replace  $\Sigma_{\text{ext}}$ with 
$\kappa(\gamma-\overline{\varepsilon})$ in Eq.\ref{eq:dym1} 
($\overline{\varepsilon}$ indicates the macroscopic strain defined as 
the spatial average of $\varepsilon(\mathbf{r})$). 
For large values of $\kappa$, it is expected that the macroscopic strain 
$\overline{\varepsilon}$ is close to $\gamma$. The slope of the stress-strain curve is given by $\mathcal{G}=2 \kappa \mu/(\kappa+2 \mu) $. Here we use $\kappa=10$ and $\mu=1/2$, so we get $10/11 \approx 0.91$ used as the initial slope of the stress-strain curve in the main text.

The energy functional defined in Eq.\ref{eq:e_plt} can be equivalently written as 
\begin{eqnarray}\label{eq:e_general}
    E_{\text{plast.}}[\varepsilon(\bm{r})] &=& \int d^2\bm{r} e(\bm{r}) \nonumber \\
    \text{with}\quad e(\bm{r}) &\hat{=}& \frac{1}{2}\varepsilon(\bm{r})\int d\bm{r}'G(\bm{r}-\bm{r}')\varepsilon(\bm{r}')  \nonumber \\
    &+& V\x\left(\bm{r},\varepsilon(\bm{r})\right) .
\end{eqnarray}
Thus our system can be viewed as an elastic manifold $\varepsilon(\bm{r})$
(with long-range interactions among all sites)
living in an energy landscape  where each site has its independent disordered 
potential. 
The stress-free energy used in the main text is then defined as the plastic energy per site:
$E_{\tt sf} \hat{=} E_{\text{plast.}}[\varepsilon*]/L^2$, with $\frac{\delta E_{\text{plast.}}}{\delta \varepsilon(\bm{r})}\big|_{\varepsilon^*}=0$. The energy profiles of the main text are obtained by averaging the local energy to $e(\bm{r})$ along the direction of the shear band. 
We apply quasi-static deformation $\gamma$ by numerically solving Eq.\ref{eq:dym1} mapped onto a lattice system. 

\begin{figure}[t!]
    \centering
     \includegraphics[width=1.\columnwidth]{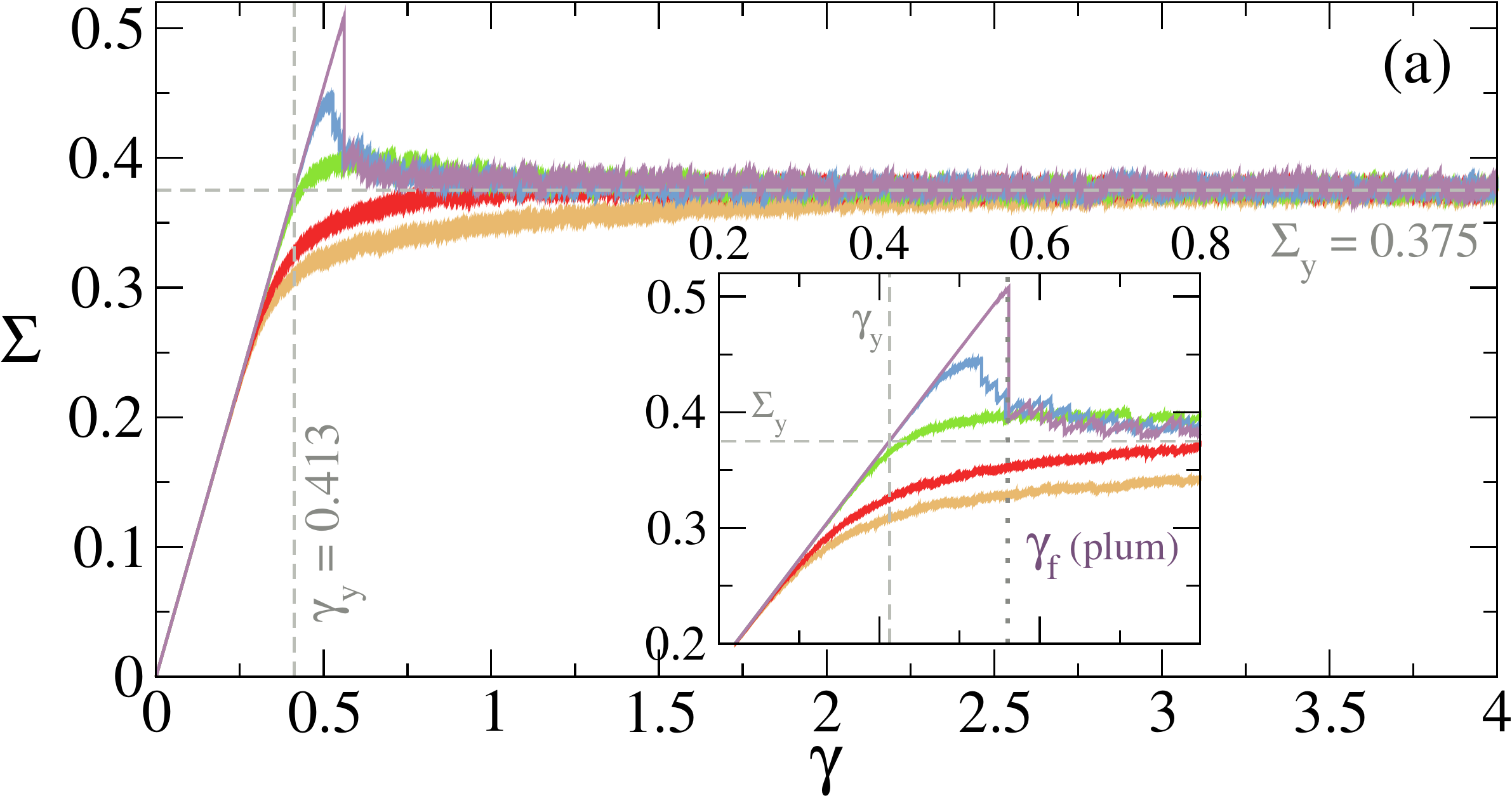}
     \includegraphics[width=1.\columnwidth]{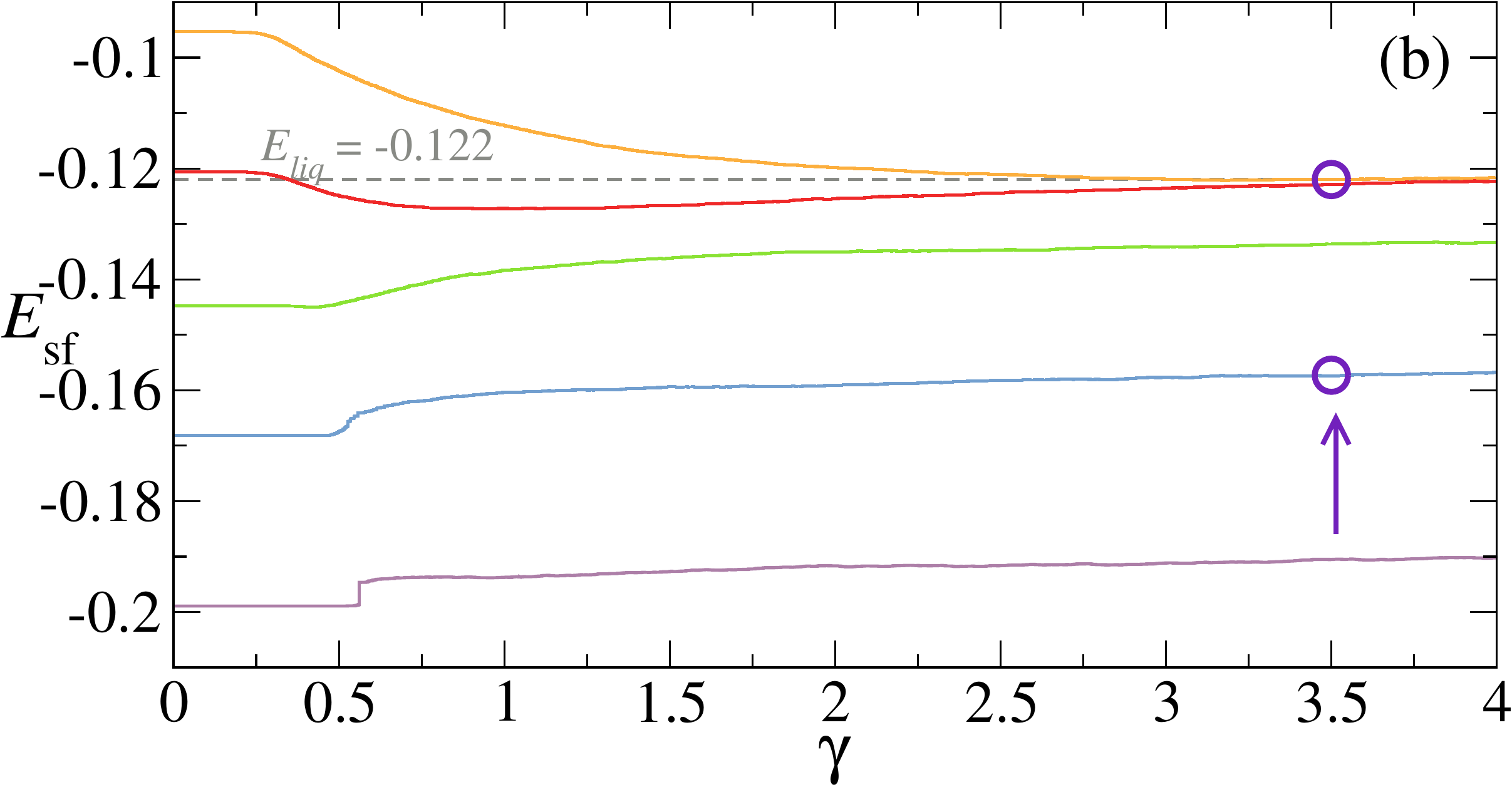}
     \includegraphics[width=0.45\columnwidth]{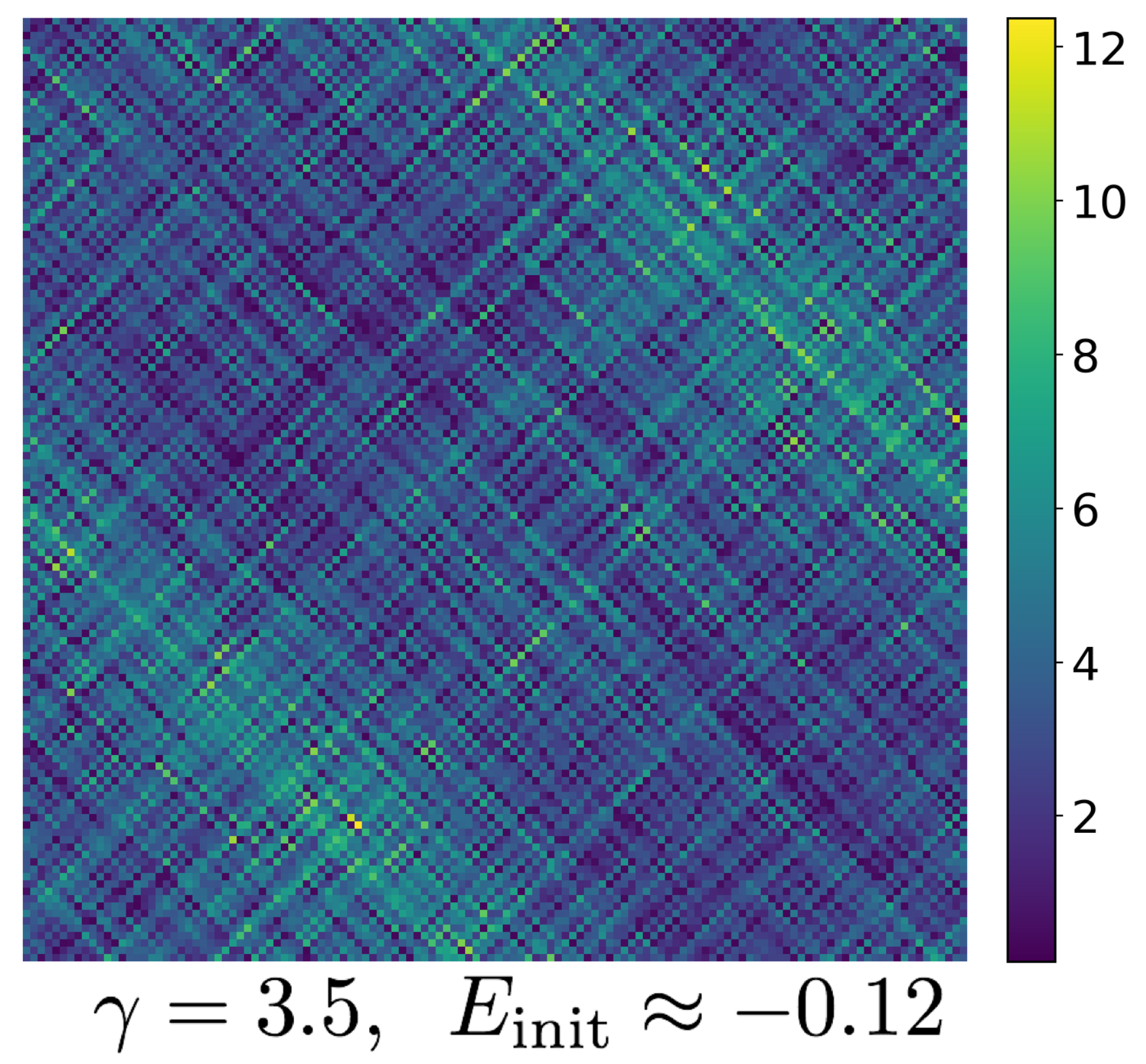}
     \includegraphics[width=0.45\columnwidth]{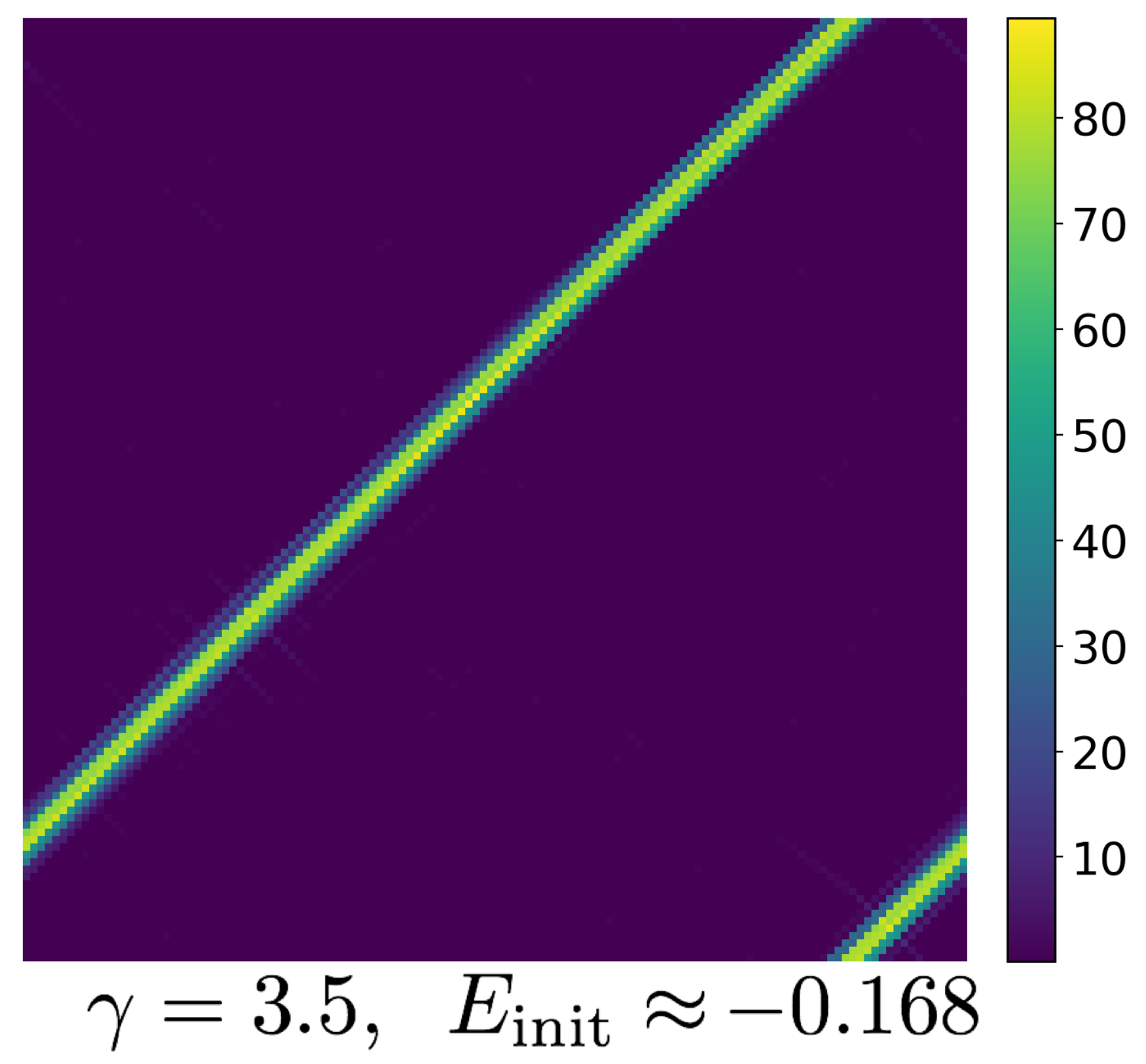}
     \put(0,80){\large(c)}
     \caption{
    Uniform shear deformation for different degrees of initial annealing 
    ($\Einit\approx$ -0.095 \text{(chocolate)}, -0.12\text{(red)}, 
    -0.149\text{(chamelon)}, -0.168\text{(blue)}, -0.199\text{(plum)}). 
    {\bf (a)} Stress-strain curves for different initial degrees of annealing. The yield stress $\Sigma_y=0.375$ defined as the plateau is also indicated.
    {\bf (b)} Evolution of the stress-free energy $E_{\tt sf}=E(\Sigma)-\Sigma^2/(4\mu)$ 
    during the deformation. The stationary plateau $E_{\text{liq}}=-0.122$ is also indicated.
    {\bf (c):}  Colormap of the local deformation field at $\gamma=3.5$ for two samples prepared at different degrees
    of annealing, circles in (b). 
    }
    \label{fig:uniformdeformation}
\end{figure}
 
Special care is taken for the initial condition, since we want to mimic
different degrees of annealing in our samples.
In our model, a well-annealed sample corresponds to an initial condition 
where the local strains are confined in deep potential wells, 
while poorly-annealed samples correspond to a strain manifold confined 
in shallow potential wells.
In practice we start from a liquid configuration of local energy basins 
obtained after a long uniform shear.
Then, to  prepare samples of different degree of annealing we change the 
depths of the parabolic wells, $\epsilon_w$.
To obtain deeply annealed samples we also align the centers, $\epsilon_{pl}$, 
to their mean value. 
Finally, we let the local strain relax to its stress-free configuration,
and that makes our initial condition to start the oscillatory deformation.
We performed simulations 
for systems of sizes $N= 32 \times 32, 64 \times 64, 128\times128, 256 \times 256$.
The data  shown in this work correspond to the two largest systems.

\section{Uniform shear deformation curves \label{Apx-uni}}

\begin{figure}[b!]
    \centering
     \includegraphics[width=0.8\columnwidth]{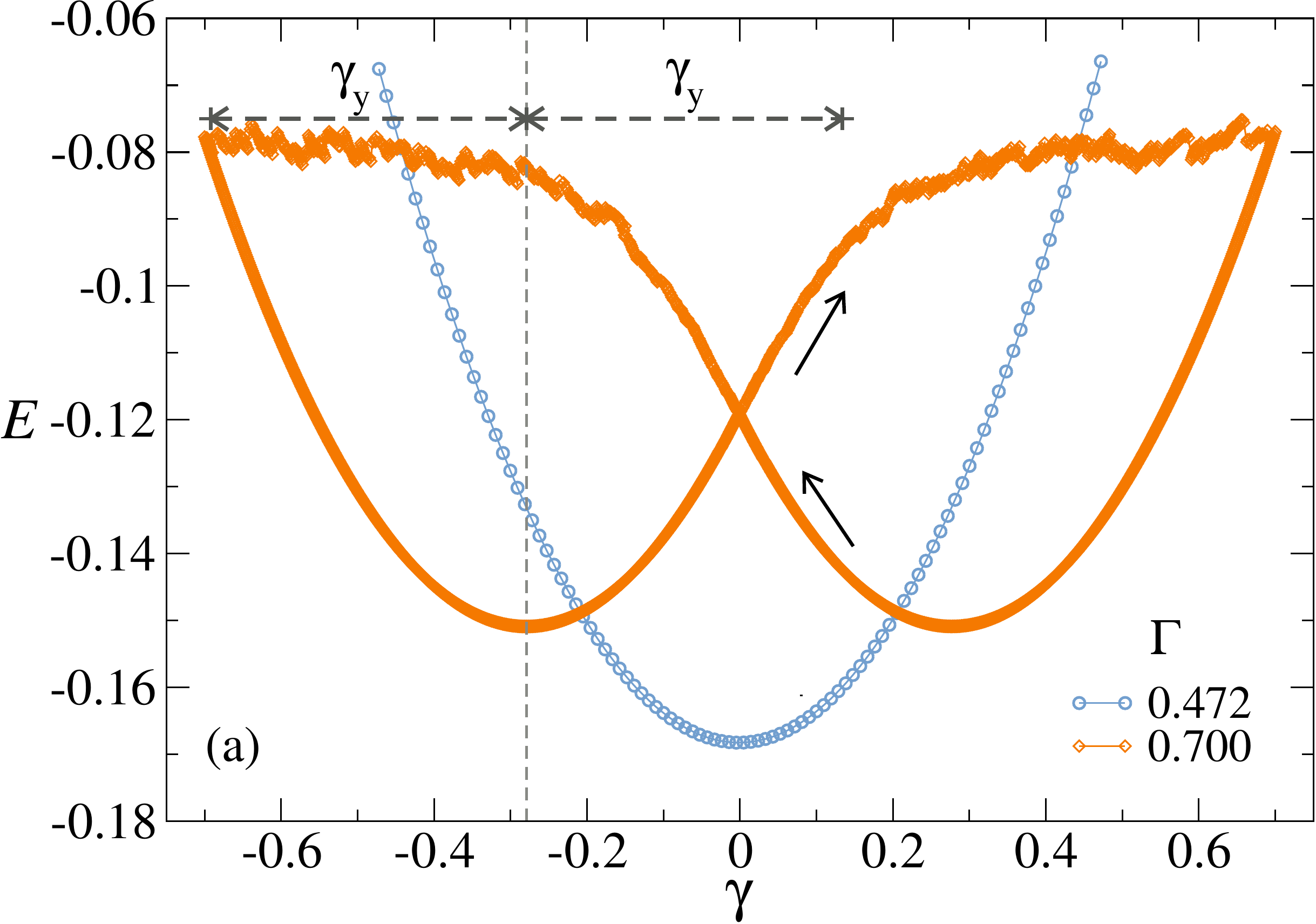}
     \includegraphics[width=0.8\columnwidth]{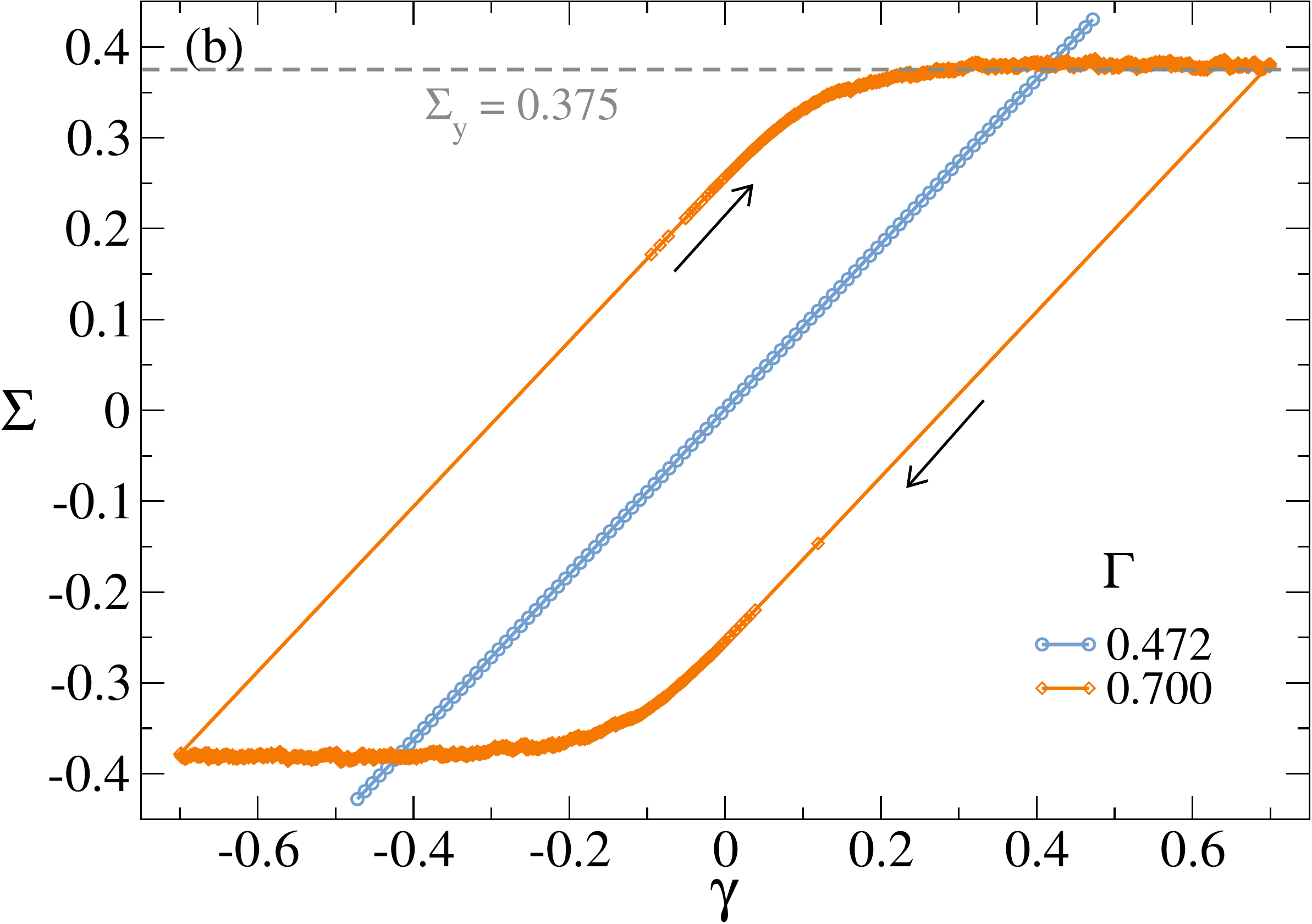}
    \caption{
    {\bf (a)} Energy versus strain in the oscillatory steady state 
    (only one loop shown) 
    for $\gmax=0.45$ (light-blue circles) and $\gmax=0.7$ (orange diamonds).
    The little black arrows mark the sense of the loop.
    {\bf (b)} Stress as a function of strain in the oscillatory steady 
    state for the same two amplitudes as above.
    }\label{fig:Loops} 
\end{figure}

We present in this section more detailed results for the case of uniform shear deformation. 
In Fig.\ref{fig:uniformdeformation}(a), we show the same stress-strain curves as Fig.1a of 
the main text with a larger scope of strain up to $\gamma=4$, so that the common plateau 
$\Sigma_y\approx 0.375(\pm 0.003)$ at long strains is well captured.
We show system's stress-free energy as function of strain in Fig.\ref{fig:uniformdeformation}(b).
Poorly annealed samples (those which do not display a failure) reach rather quickly a common 
energy plateau $\Eliq\approx -0.122(\pm0.001)$ which corresponds to the fully formed 
stationary liquid state. 
In contrast, well annealed samples display different and slowly-evolving transient regimes. 
These differences can be understood as follows:
Poorly annealed systems melt at deformations around $\gamma_y$ into a non-Newtonian 
liquid which has no memory of the initial degree of annealing 
(see Fig.\ref{fig:uniformdeformation} bottom-left).
In well annealed systems, instead, the non-Newtonian liquid is localised 
within a shear band and the solid part is unchanged, insensitive to the 
deformation (see Fig.\ref{fig:uniformdeformation} bottom-right). 
When $\gamma$ is increased further, the band widens.
As discussed in~\cite{JaglaJSTAT2010}, we expect 
$w(\gamma) \propto \sqrt{\gamma - \gamma_{\tt f}}$.
Eventually, for very large $\gamma$, the band width reaches the 
system size and the whole sample lose memory of the initial condition 
converging to the stationary value $\Eliq$. 

\section{Typical loops in oscillatory stationary states \label{Apx-loop}}
In Fig.~\ref{fig:Loops}, we show the total energy $E$ (per volume) as a function
of $\gamma$ during a steady oscillation between $\pm \gmax$.
While in the solid phase  ($\gmax<\Gamma_c$), the energy displays a single parabolic 
shape, corresponding to an elastic behavior, in the mixed phase ($\gmax>\Gamma_c$), 
the energy has a butterfly shape: 
two shifted parabola are followed by a noisy  plateau which reveals the presence of the 
liquid state.
The two parabolas have an extension of $\gamma_y$ at both sides and the maximal stress, 
at $\pm\gmax$, saturates to $\Sigma_{y}$, which also supports that the shear band is made
of the stationary liquid state at large uniform shear, where plasticity dominates the 
dynamics once sheared beyond $\gamma_y$ from a zero stress state.


\begin{figure}[t!]
    \centering
     \includegraphics[width=1.\columnwidth]{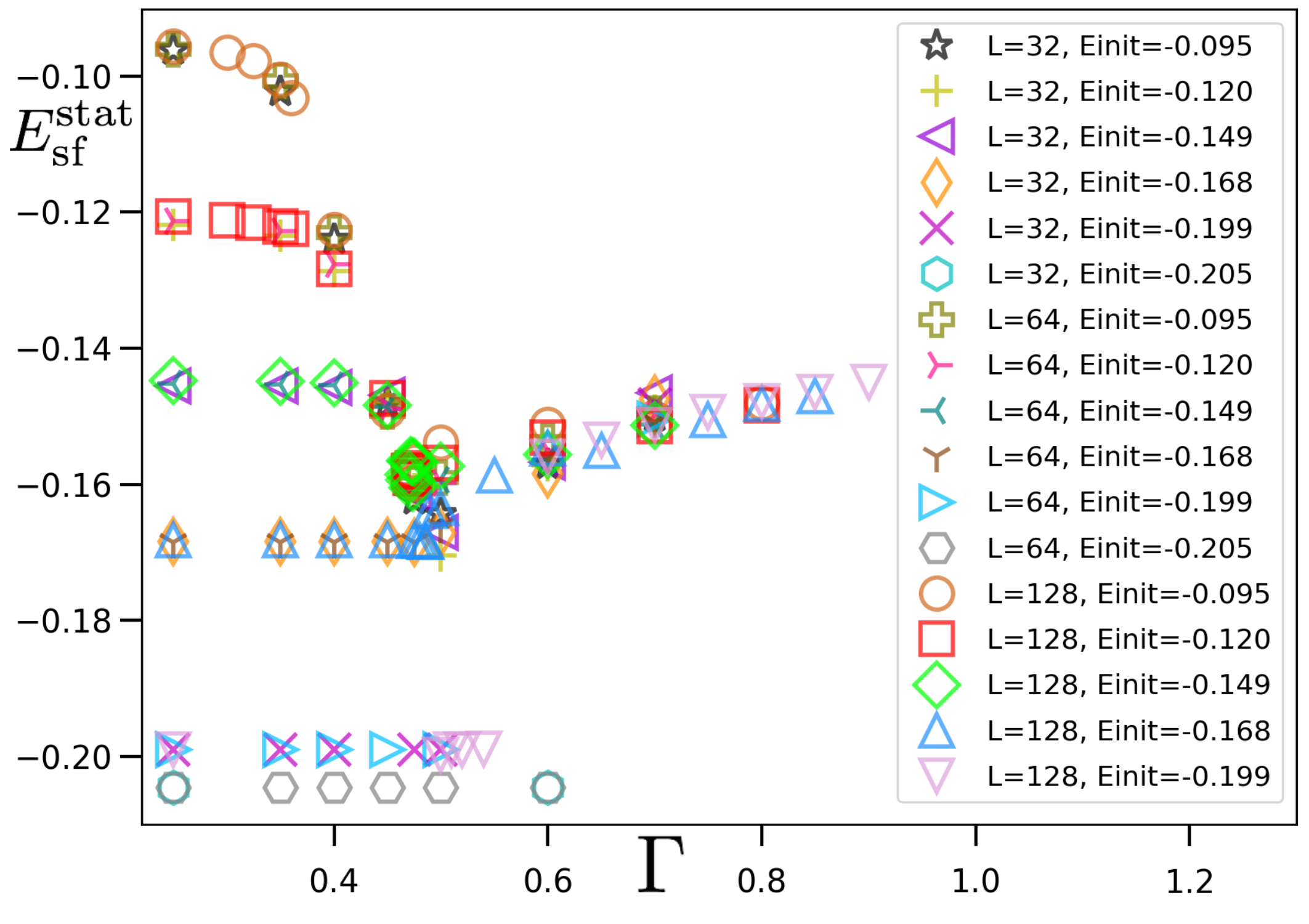}
     \caption{$E_{\tt sf}^{\tt stat}$ versus $\Gamma$ for all system sizes and initial conditions $\Einit$ as indicated in the figure. 
    }
    \label{fig:Esf_stat}
\end{figure}

\section{The discontinuous transition in $E_{\tt sf}^{\tt stat}$ across $\Gamma^*$}\label{Apx-dis}
\red{
In Fig.\ref{fig:Esf_stat} we observe that,  at different amplitudes $\Gamma$,   $E_{\tt sf}^{\tt stat}$ (namely the steady-state stress-free energies per unit volume)  does not depend on the system size. As a consequence, in the mixed phase, for  $\Gamma>\Gamma_c\geq\Gamma^*$, the shear band invades a finite fraction $r=w_s/L$ of the system, with $w_s$ representing the width of shear band. We can evaluate this fraction by observing that, in the stationary state, during the oscillation driving from  $-\Gamma$ to $\Gamma$, the material behaves initially as a solid and accumulates an elastic strain $\approx 2 \gamma_y$ (see e.g. Fig.\ref{fig:Loops}), the rest of the macroscopic deformation $2\Gamma-2\gamma_y$ consists of the plasticity confined within the shear band, denoted $\gamma_\text{plt}$. Formally, one may write: 
\begin{eqnarray}
    2\Gamma-2\gamma_y =   r \gamma_\text{plt}\;.
\end{eqnarray}
Since plasticity takes place significantly only during the loading from  $-\Gamma+2\gamma_y$ to $+\Gamma$, it is reasonable to assume
\begin{eqnarray}\label{eq:ratio}
    \gamma_\text{plt}\propto (2\Gamma-2\gamma_y)^x\;,\quad\text{with}\quad x>0.
\end{eqnarray}
Combining the two relations takes to
\begin{eqnarray}
    r(\Gamma)&=&K(\Gamma-\gamma_y)^{1-x}\;, \nonumber \\
    w_s(\Gamma) &=& r\cdot L = a(\Gamma-\gamma_y)^{1-x}\;. 
\end{eqnarray}
with $K>0$ a constant independent on the system size.
A simple guess, by taking arguments from ref.~\cite{JaglaJSTAT2010}, 
is $x\approx 1/2$, and this yields our empirically observed
$\alpha\approx 1/2$, while identifying $\Gamma_0$ with $\gamma_y$ (Fig.\ref{fig:mixture})
}

\red{
Now two scenarios are possible: (i) if $\gamma_y = \Gamma^* $, the transition is of second order in nature as the fraction of liquid material  vanishes at the transition. (ii) If  $\gamma_y < \Gamma^* $, the transition is of first order in nature as the fraction of liquid material (shear band) is finite at the transition $\Gamma^*$. We can argue that the second scenario is correct.  Indeed $\gamma_y$ is the elastic strain beyond which a liquid-state material (of $E_\text{liq}$) melts and its stress-strain curve bends~\cite{BonnRMP2017}. By our model construction, we have $|E_\text{liq}|\propto \gamma_y^2$. On the other hand, for a critical marginal solid, we have $|E^*| \propto {\Gamma^*}^2$ by its definition.  Thus, we have $\Gamma^* \propto \sqrt{|E^*|}>\sqrt{|E_\text{liq}|}\propto \gamma_y$.
}

\red{
Combining the above arguments and our empirical data analysis, we can conclude for the stress-free energy of poorly annealed samples that $E_\text{sf}^\text{stat}(\Gamma^{*-})=E^*<E^*+K(E_\text{liq}-E^*)(\Gamma^*-\gamma_y)^\alpha = E_\text{sf}^\text{stat}(\Gamma^{*+})$, i.e. a discontinuity across $\Gamma^*$, with the predicted jump ($\sim 0.0045$) appearing consistent with the fluctuation amplitude in the inset of Fig.\ref{fig:ssObserv_vs_gmax}(b).
}

\begin{figure}[t!]
    \centering
     \includegraphics[width=1.\columnwidth]{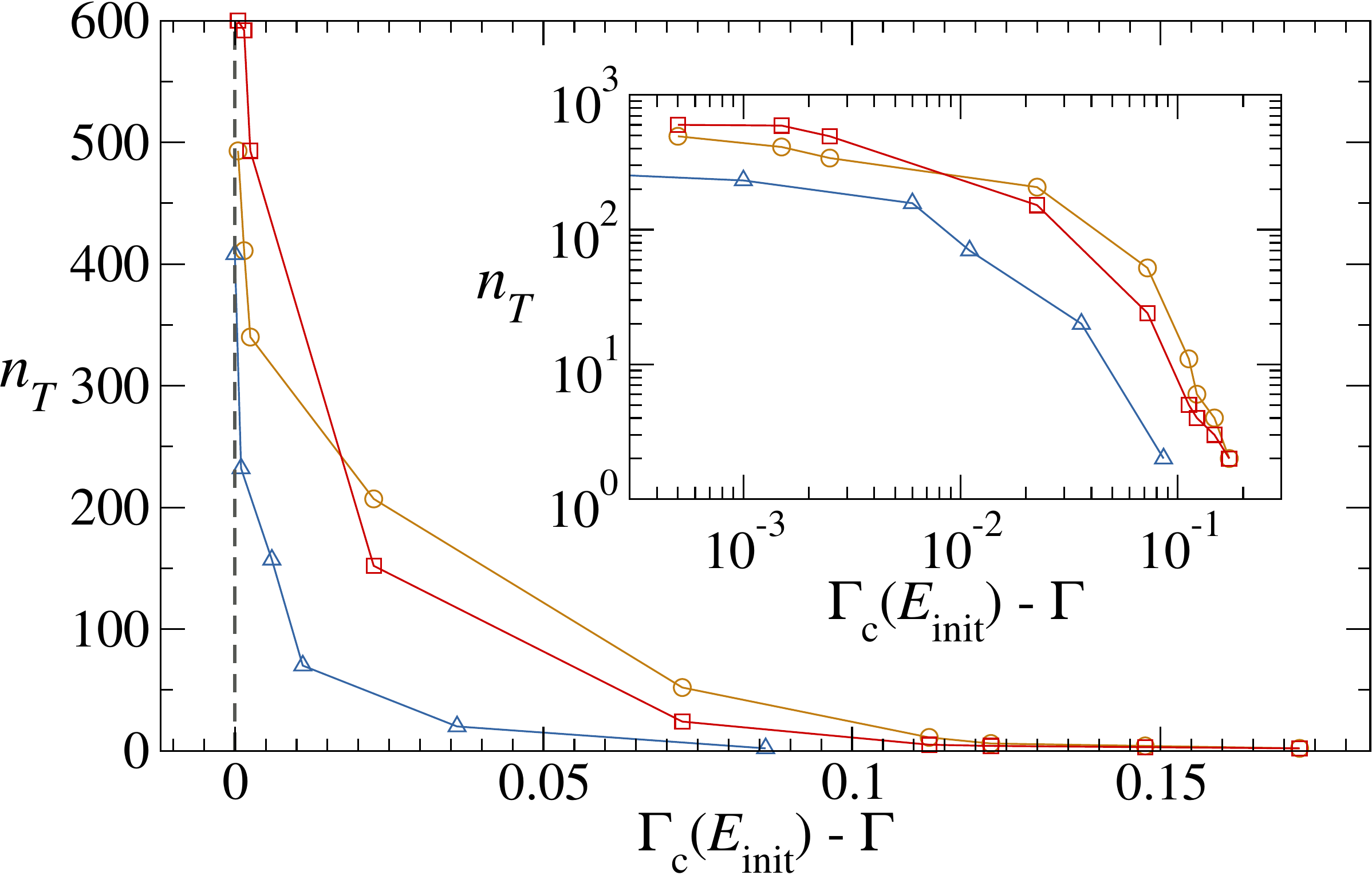}
     \caption{
    Number of cycles needed to reach the steady state as a function of the distance to the
    critical amplitude, for different initial annealing levels $\Einit$.
    }
    \label{fig:ncyclestosteady}
\end{figure}

\section{Duration of the oscillatory transient regime \label{Apx-nt}}

\red{The number of cycles needed to reach the stationary state is an observable that
that is expected to diverge at the transition, as seen equally in molecular dynamics 
of sheared amorphous solids~\cite{khirallah2021yielding,sastry2020models} and in 
models of sheared vortices in disordered media~\cite{okuma2011transition,brown2019reversible}}. 
We show in Fig.\ref{fig:ncyclestosteady} the number $n_T$ of oscillation cycles needed to 
reach the stationary state \red{in our system} for $\Gamma<\Gamma_c$ and $\Einit>E^*$. 
$n_T$ is easily measured in such situations since the stationary state is unambiguously 
identified once a perfect periodic loop of stress/energy (e.g. the blue curves in Fig.\ref{fig:Loops}) is established. 
Instead of a power-law divergence~\cite{khirallah2021yielding,sastry2020models},
we observe rather $n_T\propto\log\left(\frac{1}{\Gamma_c-\Gamma}\right)$. 
The stationary state of well annealed systems $\Einit<E^*$ oscillated at $\Gamma<\Gamma_c(\Einit)$ is strongly reminiscent of its initial condition, as mentioned in the main text, and thus reached almost immediately.
Only at $\Gamma$ very close to $\Gamma_c$, a slight increase of $n_T$ is observed due to 
increased local adjustment, the nature of which is an open question. 
Above $\Gamma_c$, when $\Gamma$ is approaching $\Gamma_c$, the system suffer a drastic 
dynamic slowing down that makes an accurate measurement of $n_T$ more difficult. 
A more careful assessment of the specific law displayed by our model for the divergence 
of $n_T$ at $\Gamma_c$ is left for future investigation.

\section{Shear-band emergence in well annealed samples \label{Apx-growth}}

\begin{figure}[t!]
\centering
\includegraphics[width=0.98\columnwidth]{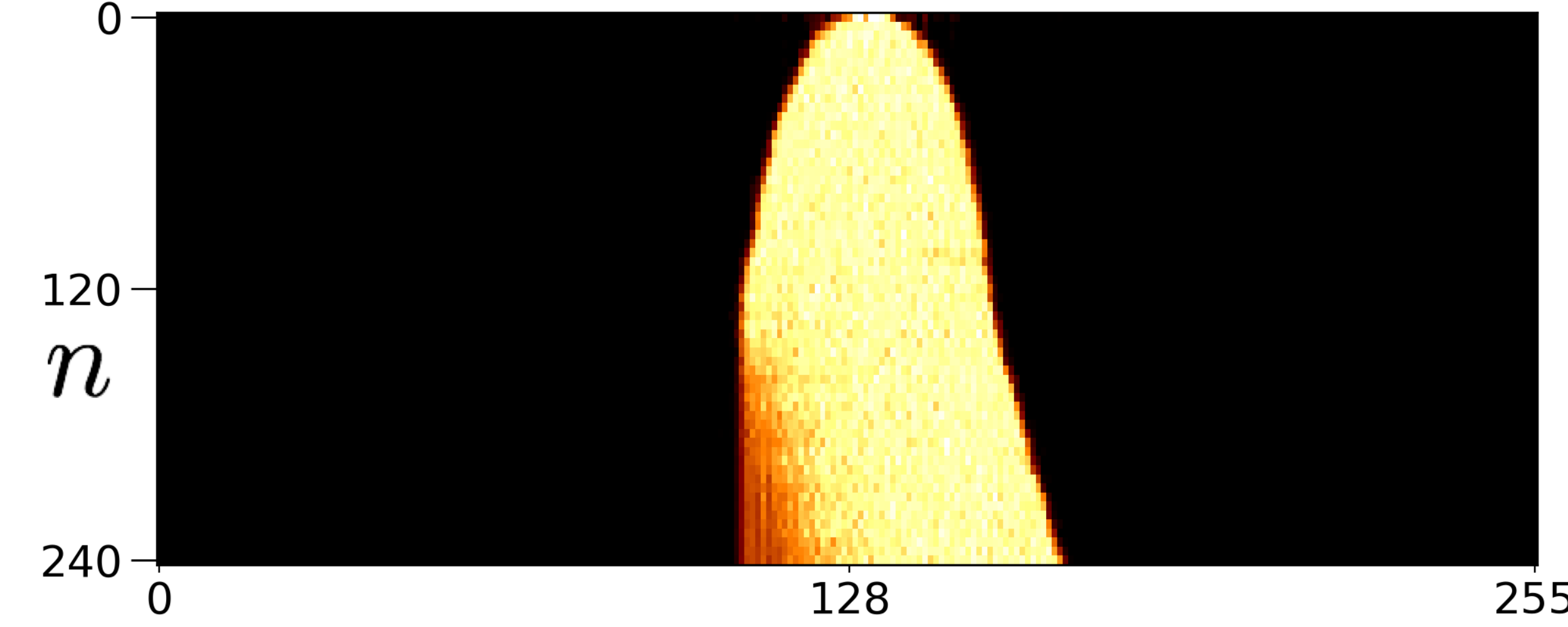}
\caption{
The energy profile across a section perpendicular to the shear band evolving as 
function of the number $n$ of oscillation cycles. 
The light color region represents the position of the shear band with a higher energy. 
The amplitude of the oscillation is $\gmax=0.7$. 
The sample, of linear  size $L=256$, has initially been prepared at $\Einit\approx -0.199$, 
i.e., in a very well annealed state.
}\label{fig:Prfl_evo_Si}
\end{figure}

In Fig.\ref{fig:Prfl_evo_Si} we show a color-map image similar to Fig.~4 of the main text.
It depicts the energy profile evolution as function of the number of oscillation cycles $n$
in the first stages of the transient dynamics of an initially well annealed sample (here a sample of system size $L=256$ for a better visualisation).
The inverted parabolic-like shape observed as $n$ increases illustrates the band {\it coarsening}.
In fact the bad-width growth law is rather $w \sim n^{1/3}$.
At about $n\sim 120$ steps, the band has reached a stable width and starts moving ballistically 
to the right, leaving behind a marginal solid.


%

\end{document}